\documentclass[12pt,letterpaper]{article}

\global\def\draftcontrol{0}

   \def\versionno{ IR Black Holes: Numerics }
\catcode`\@=11

\expandafter\ifx\csname draftcontrol\endcsname\relax\global\def\draftcontrol{0}
\fi

{\count255=\time\divide\count255 by 60
\xdef\hourmin{\number\count255}
\multiply\count255 by-60\advance\count255 by\time
\xdef\hourmin{\hourmin:\ifnum\count255<10 0\fi\the\count255}}
\def\draftdate{\number\month/\number\day/\number\year\ \ \ \hourmin }

\newcommand\makepapertitle{\par
  \begingroup
    \renewcommand\thefootnote{\@fnsymbol\c@footnote}%
    \def\@makefnmark{\rlap{\@textsuperscript{\normalfont\@thefnmark}}}%
    \long\def\@makefntext##1{\parindent 1em\noindent
            \hb@xt@1.8em{%
                \hss\@textsuperscript{\normalfont\@thefnmark}}##1}%
     \newpage
     \global\@topnum\z@   
     \@makepapertitle
     \thispagestyle{empty}\@thanks
  \endgroup
  \setcounter{footnote}{0}%
  \global\let\thanks\relax
  \global\let\makepapertitle\relax
  \global\let\@makepapertitle\relax
  \global\let\@thanks\@empty
  \global\let\@author\@empty
  \global\let\@date\@empty
  \global\let\@title\@empty
  \global\let\title\relax
  \global\let\author\relax
  \global\let\date\relax
  \global\let\and\relax
  \def\version{\let\version\@version\@gobble}
}
\def\@makepapertitle{%
  \newpage
   \ifnum\draftcontrol=1 {}
   \version\versionno
   \vskip 3em%
   \else
   \hfill\hbox to 3cm {\parbox{4cm}{\@pubnum}\hss}%
   \vskip 3em%
   \fi
   \begin{center}%
   \let \footnote \thanks
     {\LARGE {\@title}}%
     \vskip 1.5em%
     {\normalsize
       \lineskip .5em%
       \begin{tabular}[t]{c}%
         \@author
       \end{tabular}\par}%
     \vskip 1.5em%
     {\@bstract}%
     \end{center}%
     \vskip 1.5em 
     \@date%
   \par
}

\gdef\@pubnum{}
\def\pubnum#1{%
  \gdef\@pubnum{#1}}

\gdef\@bstract{}
\def\Abstract#1{%
  \gdef\@bstract{%
   \parbox{\textwidth-0pc}{%
   \centerline{\bf Abstract}\penalty1000%
\noindent
\renewcommand\baselinestretch{1.0}%
{#1}}}
}

\def\ps@paper{\let\@mkboth\@gobbletwo%
     \ifnum\draftcontrol=1
        \def\@oddfoot{\hbox to \textwidth{\tiny \versionno \hfil\tiny\draftdate}%
        \hskip -\textwidth \hbox to \textwidth{\hfil\rm\thepage\hfil}}%
     \else\def\@oddfoot{\hbox to \textwidth{\hfil\rm\thepage\hfil}}
     \fi
     \let\@evenfoot\@oddfoot
}

\def\@version#1{\ifnum\draftcontrol=1
\typeout{}\typeout{#1}\typeout{}
\vskip3mm\centerline{\hbox{\fbox{\normalsize{\tt DRAFT -- #1 -- }
                   {\draftdate}}}}\vskip3mm
\fi}
\let\version\@version
\long\def\eqlabel#1{\ifnum\draftcontrol=1
                    \tag@false  
                    \tag*{(\theequation) \hbox to -0.2cm{\hspace{0cm}\small{#1}\hss}}
                    \refstepcounter{equation} 
                    \edef\@currentlabel{\theequation}
                    \ltx@label{#1}          
                    \else
                    \label{#1}
                    \fi
                    }
\let\st@bibitem\@bibitem
\let\st@lbibitem\@lbibitem
\ifnum\draftcontrol=1
  \def\@bibitem#1{%
    \st@bibitem{#1}\a@@label{#1}\ignorespaces}
  \def\@lbibitem[#1]#2{%
    \st@lbibitem[#1]{#2}\a@@label{#2}\ignorespaces}
  \def\a@@label#1{%
    \gdef\a@lab{\smash{\normalfont\small#1}}
    \ifvmode
      \if@inlabel
        \global\setbox\@labels\hbox{%
          \llap{\a@lab\let\a@lab\relax
                \kern\@totalleftmargin\kern\marginparsep}%
          \box\@labels}%
      \fi
    \fi}
\fi

\usepackage{amscd}
\usepackage{graphicx} 
\usepackage{amsmath,amssymb,array,calc,rotating}

\ifnum\draftcontrol=1
\fi

\renewcommand\baselinestretch{1.25}
\renewcommand\section{\@startsection {section}{1}{\z@}%
                                   {-3.5ex \@plus -1ex \@minus -.2ex}%
                                   {2.3ex \@plus.2ex}%
                                   {\normalfont\large\bfseries}}
\renewcommand\subsection{\@startsection{subsection}{2}{\z@}%
                                   {-3.25ex\@plus -1ex \@minus -.2ex}%
                                   {1.5ex \@plus .2ex}%
                                   {\normalfont\normalsize\bfseries}}
\renewcommand\subsubsection{\@startsection{subsubsection}{3}{\z@}%
                                   {-3.25ex\@plus -1ex \@minus -.2ex}%
                                   {1.5ex \@plus .2ex}%
                                   {\normalfont\normalsize\it}}
\renewcommand\paragraph{\@startsection{paragraph}{4}{\z@}%
                                   {-3.25ex\@plus -1ex \@minus -.2ex}%
                                   {1.5ex \@plus .2ex}%
                                   {\normalfont\normalsize\bf}}

\def\revise#1       {\raisebox{-0em}{\rule{3pt}{1em}}%
                     \marginpar{\raisebox{.5em}{\vrule width3pt\
                     \vrule width0pt height 0pt depth0.5em
                     \hbox to 0cm{\hspace{0cm}{%
                     \parbox[t]{4em}{\raggedright\footnotesize{#1}}}\hss}}}}

\def\call         {{\cal L}}

\def\del          {\partial}

\def\ee           {{\rm e}}

\def\tr           {\mathop{\rm Tr}}

\def\de#1#2{{\rm d}^{#1}\!#2\,}

\def\sqr#1#2{{\vcenter{\vbox{\hrule height.#2pt  
 \hbox{\vrule width.#2pt height#1pt \kern#1pt
 \vrule width.#2pt}\hrule height.#2pt}}}}

\def\a{\alpha}
\def\b{\beta}
\def\r{\rho}

\def\la{\lambda}

\def\be{\begin{equation}}
\def\ee{\end{equation}}

\def\m{\mu}
\def\g{\gamma}
\def\l{\lambda}
\def\n{\nu}

\catcode`\@=12

\begin{document}


   
\def \el {{\ell}}   
\def \KK {{\cal  K}}   
\def \K {{\rm K}}   
\def \tz{\tilde{z}}   
   
\def \ci {\cite}   
\newcommand{\rf}[1]{(\ref{#1})}   
\def \la {\label}   
\def \const {{\rm const}}

\def \ov {\over}   
\def \ha {\textstyle { 1\ov 2}}   
\def \we { \wedge}   
\def \P { \Phi} \def\ep {\epsilon}   
\def \ab {{A^2 \ov B^2}}   
\def \ba {{B^2 \ov A^2}}   
\def \tv   {{1 \ov 12}}   
\def \go { g_1}\def \gd { g_2}\def \gt { g_3}   
\def \gc { g_4}\def \gp { g_5}\def \F {{\cal F}}   
\def \del { \partial}   
\def \t {\theta}   
\def \p {\phi}   
\def \ep {\epsilon}   
\def \te {\tilde \epsilon}   
\def \ps {\psi}   
\def \x {{x_{11}}}

\def\br{\bar{\rho}}   
\newcounter{subequation}[equation]

\def\pa{\partial}   
\def\e{\epsilon}   
\def\rt{\rightarrow}   
\def\tr{{\tilde\rho}}   
\newcommand{\eel}[1]{\label{#1}\end{equation}}   
\newcommand{\bea}{\begin{eqnarray}}   
\newcommand{\eea}{\end{eqnarray}}   
\newcommand{\eeal}[1]{\label{#1}\end{eqnarray}}   
\newcommand{\LL}{e^{2\lambda(r)}}   
\newcommand{\NN}{e^{2\nu(r)}}   
\newcommand{\PP}{e^{-2\phi(r)}}   
\newcommand{\non}{\nonumber \\}   
\newcommand{\CR}{\non\cr}

\makeatletter   
   
\def\thesubequation{\theequation\@alph\c@subequation}   
\def\@subeqnnum{{\rm (\thesubequation)}}   
\def\slabel#1{\@bsphack\if@filesw {\let\thepage\relax   
   \xdef\@gtempa{\write\@auxout{\string   
      \newlabel{#1}{{\thesubequation}{\thepage}}}}}\@gtempa   
   \if@nobreak \ifvmode\nobreak\fi\fi\fi\@esphack}   
\def\subeqnarray{\stepcounter{equation}   
\let\@currentlabel=\theequation\global\c@subequation\@ne   
\global\@eqnswtrue \global\@eqcnt\z@\tabskip\@centering\let\\=\@subeqncr   
   
$$\halign to \displaywidth\bgroup\@eqnsel\hskip\@centering   
  $\displaystyle\tabskip\z@{##}$&\global\@eqcnt\@ne   
  \hskip 2\arraycolsep \hfil${##}$\hfil   
  &\global\@eqcnt\tw@ \hskip 2\arraycolsep   
  $\displaystyle\tabskip\z@{##}$\hfil   
   \tabskip\@centering&\llap{##}\tabskip\z@\cr}   
\def\endsubeqnarray{\@@subeqncr\egroup   
                     $$\global\@ignoretrue}   
\def\@subeqncr{{\ifnum0=`}\fi\@ifstar{\global\@eqpen\@M   
    \@ysubeqncr}{\global\@eqpen\interdisplaylinepenalty \@ysubeqncr}}   
\def\@ysubeqncr{\@ifnextchar [{\@xsubeqncr}{\@xsubeqncr[\z@]}}   
\def\@xsubeqncr[#1]{\ifnum0=`{\fi}\@@subeqncr   
   \noalign{\penalty\@eqpen\vskip\jot\vskip #1\relax}}   
\def\@@subeqncr{\let\@tempa\relax   
    \ifcase\@eqcnt \def\@tempa{& & &}\or \def\@tempa{& &}   
      \else \def\@tempa{&}\fi   
     \@tempa \if@eqnsw\@subeqnnum\refstepcounter{subequation}\fi   
     \global\@eqnswtrue\global\@eqcnt\z@\cr}   
\let\@ssubeqncr=\@subeqncr   
\@namedef{subeqnarray*}{\def\@subeqncr{\nonumber\@ssubeqncr}\subeqnarray}   
   
\@namedef{endsubeqnarray*}{\global\advance\c@equation\m@ne   
                           \nonumber\endsubeqnarray}   
   
\makeatletter \@addtoreset{equation}{section} \makeatother   
\renewcommand{\theequation}{\thesection.\arabic{equation}}   
   
\def \ci {\cite}   
\def \la {\label}   
\def \const {{\rm const}}   
\catcode`\@=11   
   
\newcount\hour   
\newcount\minute   
\newtoks\amorpm \hour=\time\divide\hour by 60\minute   
=\time{\multiply\hour by 60 \global\advance\minute by-\hour}   
\edef\standardtime{{\ifnum\hour<12 \global\amorpm={am}   
        \else\global\amorpm={pm}\advance\hour by-12 \fi   
        \ifnum\hour=0 \hour=12 \fi   
        \number\hour:\ifnum\minute<10   
        0\fi\number\minute\the\amorpm}}   
\edef\militarytime{\number\hour:\ifnum\minute<10 0\fi\number\minute}   
   
\def\draftlabel#1{{\@bsphack\if@filesw {\let\thepage\relax   
   \xdef\@gtempa{\write\@auxout{\string   
      \newlabel{#1}{{\@currentlabel}{\thepage}}}}}\@gtempa   
   \if@nobreak \ifvmode\nobreak\fi\fi\fi\@esphack}   
        \gdef\@eqnlabel{#1}}   
\def\@eqnlabel{}   
\def\@vacuum{}   
\def\marginnote#1{}   
\def\draftmarginnote#1{\marginpar{\raggedright\scriptsize\tt#1}}   
\overfullrule=0pt   
   
 \def \lc {light-cone\ }   
   
\def\draft{   
        \pagestyle{plain}   
        \overfullrule=2pt   
        \oddsidemargin -.5truein   
        \def\@oddhead{\sl \phantom{\today\quad\militarytime} \hfil   
        \smash{\Large\sl DRAFT} \hfil \today\quad\militarytime}   
        \let\@evenhead\@oddhead   
        \let\label=\draftlabel   
        \let\marginnote=\draftmarginnote   
        \def\ps@empty{\let\@mkboth\@gobbletwo   
        \def\@oddfoot{\hfil \smash{\Large\sl DRAFT} \hfil}   
        \let\@evenfoot\@oddhead}   
   
\def\@eqnnum{(\theequation)\rlap{\kern\marginparsep\tt\@eqnlabel}   
        \global\let\@eqnlabel\@vacuum}  }   
   
\renewcommand{\rf}[1]{(\ref{#1})}   
\renewcommand{\theequation}{\thesection.\arabic{equation}}   
\renewcommand{\thefootnote}{\fnsymbol{footnote}}   
   
\newcommand{\newsection}{    
\setcounter{equation}{0}   
\section}   
   
\textheight = 22truecm    
\textwidth = 17truecm    
\hoffset = -1.3truecm    
\voffset =-1truecm   
   
\def \tx {\textstyle}   
\def \tix{\tilde{x}}   
\def \bi{\bibitem}   
   
\def \ov {\over}   
\def \ha {\textstyle { 1\ov 2}}   
\def \we { \wedge}   
\def \P { \Phi} \def\ep {\epsilon}   
\def \ab {{A^2 \ov B^2}}   
\def \ba {{B^2 \ov A^2}}   
\def \tv   {{1 \ov 12}}   
\def \go { g_1}\def \gd { g_2}\def \gt { g_3}   
\def \gc { g_4}\def \gp {   
g_5}   
\def \F {{\cal F}}   
\def \del { \partial}   
\def \t {\theta}   
\def \p {\phi}   
\def \ep {\epsilon}   
\def \ps {\psi}

\def \LL{{\cal L}}   
\def\o{\omega}   
\def\O{\Omega}   
\def\e{\epsilon}   
\def\pd{\partial}   
\def\pdz{\partial_{\bar{z}}}   
\def\bz{\bar{z}}   
\def\e{\epsilon}   
\def\m{\mu}   
\def\n{\nu}   
\def\a{\alpha}   
\def\b{\beta}   
\def\g{\gamma}   
\def\G{\Gamma}   
\def\d{\delta}   
\def\r{\rho}   
\def\bx{\bar{x}}   
\def\by{\bar{y}}   
\def\bm{\bar{m}}   
\def\bn{\bar{n}}   
\def\s{\sigma}   
\def\na{\nabla}   
\def\D{\Delta}   
\def\l{\lambda}   
\def\te{\theta} \def \t {\theta}   
\def\ta {\tau}   
\def\na{\bigtriangledown}   
\def\p{\phi}   
\def\L{\Lambda}   
\def\hR{\hat R}   
\def\ch{{\cal H}}   
\def\ep{\epsilon}   
\def\bj{\bar{J}}   
\def \foot{ \footnote}   
\def\be{\begin{equation}}   
\def\ee{\end{equation}}   
\def \P {\Phi}   
\def\un{\underline{n}}   
\def\ur{\underline{r}}   
\def\um{\underline{m}}   
\def \ci {\cite}   
\def \g {\gamma}   
\def \G {\Gamma}   
\def \k {\kappa}   
\def \l {\lambda}   
\def \L {{L}}   
\def \Tr {{\rm Tr}}   
\def\apr{{A'}}   
\def \m {\mu}   
\def \n {\nu}   
\def \W{{\cal W}}   
\def \eps {\epsilon}   
\def \ha{{   
 { 1 \ov 2}} }   
\def \de{{   
{ 1 \ov 9}} }   
\def \si{{   
 { 1 \ov 6}} }   
\def \fo{{   
{ 1 \ov 4}} }   
\def \ei{{   
{ 1 \ov 8}} }   
\def \rt {{\tx { \ta \ov 2}}}   
\def \rr {{\bar \rho}}   
   
\def\D{\Delta}   
\def\l{\lambda}   
\def\L{\Lambda}   
\def\te{\theta}   
\def\g{\gamma}   
\def\Te{\Theta}   
\def\tw{\tilde{w}}

\def\sn{\rm sn} 
\def\cn{\rm cn}  
\def\dn{\rm dn}

\def\hzero{\hat{0}}
\def\ha{\hat{a}}
\def\hb{\hat{b}}
\def\hc{\hat{c}}
\def\hd{\hat{d}}
\def\he{\hat{e}}

\def\hone{\hat{1}}
\def\htwo{\hat{2}}
\def\hthree{\hat{3}}
\def\hz{\hat{z}}
\def\hteone{\hat{\theta}_1}
\def\htetwo{\hat{\theta}_2}
\def\hpone{\hat{\phi}_1}
\def\hptwo{\hat{\phi}_2}
\def\hpsi{\hat{\psi}}


\topmargin=0.50in  
   
\date{}   
   
\begin{titlepage}   
   
\version\versionno  
   
\hfill hep-th/0605170   
   
\hfill MCTP-06-08\\   
   
\begin{center}   
   
{\Large\bf Black Holes with Varying Flux:}\\

\vskip .7 cm 
{\Large \bf  A Numerical Approach}

\vskip .3cm 

\vskip .7 cm   
{\Large Leopoldo A. Pando Zayas$^{1}$ and C\'esar A. Terrero-Escalante$^{2}$}
\vskip 1 cm

\end{center}

\vskip .4cm 
\centerline{\it ${}^1$ Michigan Center for Theoretical Physics}   
\centerline{ \it Randall Laboratory of Physics, The University of Michigan}   
\centerline{\it Ann Arbor, MI 48109-1120, USA}    
   
\vskip .4cm 

\centerline{\it ${}^2$  Departamento de F\'isica Te\'orica, Instituto de F\' isica}   
\centerline{ \it Universidade do Estado do Rio de Janeiro}   
\centerline{\it Maracan\~a, 20559-900 RJ, Brazil }


\vskip 1.5 cm   
   
\begin{abstract}   
We present a numerical study of type IIB supergravity solutions with varying Ramond-Ramond flux. 
We construct solutions that have a regular horizon and contain nontrivial five- and three-form fluxes. 
These solutions are holographically dual to the deconfined phase of confining field theories at finite temperature. 
As a calibration of the numerical method we first numerically reproduce various analytically 
known solutions including singular and regular nonextremal D3 branes, 
the Klebanov-Tseytlin solution and its singular nonextremal generalization. 
The horizon of the solutions we construct is of the precise form of nonextremal D3 branes. 
In the asymptotic region far away from the horizon we observe a logarithmic behavior similar to that of the 
Klebanov-Tseytlin solution. 
\end{abstract}

\end{titlepage}   
\setcounter{page}{1} \renewcommand{\thefootnote}{\arabic{footnote}}   
\setcounter{footnote}{0}   
   
\def \N{{\cal N}}    
\def \ov {\over}

\section{Introduction}  

The AdS/CFT correspondence provides an alternative way to study the strongly coupled regime of gauge theories
via  gravity duals. The original statement of the AdS/CFT correspondence identifies ${\cal N}=4$ supersymmetric Yang-Mills with 
IIB strings on $AdS_5\times S^5$ \cite{malda,agmoo}. At finite temperature the gravity side is described by nonextremal D3 branes and 
the qualitative matching of the properties was one of the key observations in the understanding and eventual formulation 
of the AdS/CFT correspondence \cite{igortemp}. Other interesting aspects of finite temperature theories, as seen 
by the AdS/CFT correspondence  were 
discussed by Witten \cite{wittenhp}. In particular, the Hawking-Page phase transition in the gravity side was related to
the confinement/deconfinement transition on the field theory side. 

In a series of papers, Klebanov and collaborators 
\cite{kw,gk,kn,kt,ks,hko} developed a gravity theory that encodes 
interesting field theory phenomena like chiral symmetry breaking and 
confinement. The class of Klebanov-Tseytlin (KT) solutions are 
solutions of IIB supergravity with nontrivial metric and $F_3, H_3$ 
and $F_5$ forms. A phenomenologically very attractive property of 
the supergravity solution is that it encodes the logarithmic running 
of a gauge coupling in field theory. It does so via a varying $B_2$ 
field which is compensated by a constant $F_3$ flux through a 
3-cycle.  The 
five-form, which is constant in most solutions, varies according to the Bianchi identity 
$dF_5=H_3\wedge F_3$ and generates a varying flux as 
$\int_{\Sigma_5} F_5$ depends on the radial coordinate. The supergravity solution therefore 
has varying flux.

The finite temperature phase of such theory is certainly very 
interesting and has been tackled in various papers including 
\cite{alex,105,172}. In particular, reference \cite{172} constructed 
a perturbative solution whose regime of validity is restricted to 
high temperature and small value of the $F_3$ flux: $\int_{\Sigma_3} 
F_3 =P$. Knowing the solution only asymptotically in the radial 
coordinate and for a specific regime of parameters prevents us from 
extracting the full thermodynamics and from being able to understand possible phase transitions. 
{\it In this paper we construct numerical solutions with regular 
finite-area horizons and nonvanishing values of the five and three-form fluxes}. 
In general we find that the solutions have horizons of the form of the nonextremal D3 branes and in the other 
asymptotic region are particular generalizations of the KT class of solutions with the expected logarithmic behavior of the 
fields.

Although we have motivated the study of these backgrounds from the AdS/CFT point of view it is worth 
mentioning that from the purely gravitational point of view these are novel backgrounds. Black holes 
with varying flux have not been studied outside the context of AdS/CFT. These black holes pose 
interesting problems in the gravitational sense. For example, the definition of conserved charges is 
ambiguous \cite{marolf}. Recently, a better understanding of their thermodynamics and other properties was 
provided in \cite{aby,a2}. Our study might also be of interest to numerical relativists given that the qualitative 
properties of the solution strongly depend on the parameters and initial conditions. For instance, we 
intuitively expect the existence of the horizon to depend on various parameters including the temperature itself. 
For very low temperature we 
expect that the solution will either develop singularities or be suppressed with respect to the Klebanov-Strassler background.
Numerical methods are particularly suited to tackled these kind of problems. 

Strictly speaking, our result is relevant for a finite temperature field theory that 
has ${\cal N}=1$ supersymmetry at zero temperature. Interestingly, this theory is confining and 
we expect that
this line of research  might eventually provide a framework for understanding RHIC 
physics. 

The organization of the paper is as follows. In section \ref{sec:review} we present the Ansatz for the 
metric and form fields involved in the solution, we also reduce the ten-dimensional problem to a system of equations depending 
only on the radial coordinate. Section \ref{sec:review} also contains the analytic form of various known solutions. In 
section \ref{sec:numerics} we present a description of the numerical method and study numerically the known solutions 
presented in section \ref{sec:review}. The error of the numerical approach is estimated as the square root of the 
square difference between the known analytical solutions and our numerical solutions. We establish that 
typically this error is $\chi< 10^{-13}$.  
Section \ref{sec:new} contains our main result which is a numerical solution with 
regular horizon of finite area and three-form flux turned on top of the standard five-form. We conclude 
in section \ref{conclusions}.  In appendix \ref{geodesics} we discussed the geodesic behavior of various typical solutions to show 
how these features can be used to identify properties of the new solution.

\section{Review of gravity backgrounds with varying flux}\label{sec:review}
In this section we describe the setup from the  ten-dimensional 
point of view and reduce it to a system of ordinary differential equations. The work presented in this 
section is  largely a review of \cite{105,172} and we refer the reader to those papers for the original 
presentation and details. Our intention is to be self-contained  since we are going to use these analytical 
solutions to calibrate the numerical method.

The main idea in constructing the solution is to replace $S^5$ by a five dimensional manifold known as  $T^{1,1}$
which is topologically a product of a 2- and a 3-sphere. This 
manifold will be parametrized by coordinates $(\psi, \theta_1, \phi_1,\theta_2,\phi_2)$. 
The Ansatz for a nonzero temperature generalization of the KT solution contains 
various fields. The construction of the Ansatz follows directly the one presented  originally in 
\cite{105}. 
For the metric we consider a generalization consistent with the $U(1)$ symmetry
generated by $\psi$-rotations.\footnote{In the gauge theory \cite{kw} this symmetry is identified with 
the $U(1)_R$. Restoring this symmetry at high temperature is understood as chiral symmetry restoration \cite{172}.} 
The Ansatz in question depends on four 
functions  $x,y,z,w$   of the radial coordinate denoted by $u$:

\be
\eqlabel{eq:metric}
ds^2 =  e^{2z} ( -e^{-6x} dX_0^2 + e^{2x} dX_i dX^i)
+ e^{-2z}  ds^2_6 \ , 
\ee
where
\bea
ds^2_6 &= & e^{10y} du^2 + e^{2y} (dM_5)^2  \ ,  \nonumber \\
(dM_5)^2 & = & e^{ -8w}  e_{\psi}^2 +  e^{ 2w}
\big(e_{\theta_1}^2+e_{\phi_1}^2 +
e_{\theta_2}^2+e_{\phi_2}^2\big) \equiv
 e^{ 2w} ds_5^2.
\eea
The Funfbein is:
\be
 e_{\psi} =  {\frac13} (d\psi +  \cos \theta_1 d\phi_1  +  \cos \theta_2 d\phi_2)  \  , 
 \quad  e_{\theta_i}=\frac1{\sqrt 6} d\theta_i\ ,  \quad  e_{\phi_i}=
\frac1{\sqrt 6} \sin\theta_id\phi_i \ .
\ee
The coordinate $X^0$ represents time and  $X^i$ are the 3 longitudinal directions of the 3-brane.
The qualitative meaning of the metric functions $x,y,w$ and $z$ can be clarified from the Ansatz. The function
$x$ breaks the Poincare invariance in the four plane defined by $(X^0,X^i)$ and therefore describes the 
nonextremality. The function $z$ multiplies the four-dimensional and the six-dimensional subspaces of the metric 
by different factors and can be clearly interpreted as the warp factor. The function $y$ basically amounts to a choice of the 
parametrization of the radial coordinate, it introduces a natural ambiguity that 
we discuss later on. Finally, the function $w$ describes how the $U(1)$ fiber is squashed 
with respect to the four-dimensional base. 

To reinforce the meaning of the metric functions $x,y,z$ and $w$, it is convenient to use that
the intuition for these gravity solutions has been developed in the context of 
D3 branes. We recall that the relation with D3 branes can be established as follows:
\be
\eqlabel{eq:d3}
ds^2 =  h^{-1/2}(\r)  [ -g_1(\r)  dX_0^2 + dX_i dX_i]
+  h^{1/2}(\r)   [  g^{-1}_2 (\r) d\r^2 
+ \r^2  ds^2_5] \ , 
\ee 
with the identifications 
\be
 h=  e^{-4z- 4x} \ , \ \ \ \ \     \r  = e^{y + x + w  }\ ,
 \ \ \
\ \ \     g_1=  e^{-8x}\  ,\ \ \ \ \ \ \ e^{10y + 2x} du^2  = g_2^{-1} (\r) d\r^2 \ .  
\ee
Note that in the absence of nonextremality $(x=0)$, $h=e^{4z}$ which shows that $z$ is truly the warp factor. 
In the absence of $U(1)$ fiber squashing, that is  $w=0$, one has $e^{4y}=\rho^4= \frac1{4u} $,
the transverse 6-d space is the standard conifold
with $M_5= T^{1,1}$; this shows that $y$ amounts to a choice of the radial coordinate.  
Small  values of $u$  correspond to large distances in $\rho$ and vice versa.

The Ansatz for the $p$-form  fields is dictated by 
symmetries  and will be taken to be as in the original KT 
solution \cite{kt}:

\bea
F_3 &=& \   P
e_\psi \wedge
( e_{\theta_1} \wedge e_{\phi_1} - 
e_{\theta_2} \wedge e_{\phi_2})\ , \nonumber \\ 
B_2  &=& \    f(u) 
( e_{\theta_1} \wedge e_{\phi_1} - 
e_{\theta_2} \wedge e_{\phi_2})
 \ , \nonumber \\
F_5&=& {\cal F}+*
{\cal F}\
, \quad  \ \ \ \ {\cal F} = 
K(u) 
e_{\psi}\wedge e_{\te_1} \wedge
e_{\p_1} \wedge
e_{\te_2}\wedge e_{\p_2}\ . 
\eea
Note that the form of $F_3$ is such that it describes a 
constant flux along a 3-cycle, that is, $\int_{\Sigma_3}F_3=P$.  In some other notation this is called the number of 
fractional D3 branes. As briefly alluded to in the introduction, the  Bianchi identity for the
5-form, \ $
d*F_5=dF_5=H_{3}\wedge F_3$,  implies
\be
K (u)  = Q + 2 P f (u) \ .
\ee
That is, in the presence of 3-form flux $(P\ne 0)$, the flux of $F_5$ varies with the radius. 
The fact that $K(u)$ depends on the coordinate $u$ is very novel and has 
interesting physical implications. 

One can summarize the presence of flux in the background by noticing that all 
$p$-form fields are completely determined by a constant $P$ and a function of the radial coordinate
 $f(u)$.

\subsection{The system}
The system of equations was obtained in \cite{105} by reducing the problem to 
 a one dimensional effective 
action for the radial evolution. We refer the reader to \cite{105} for details of the derivation. The  idea 
is to plug the metric and all the forms in an effective action for IIB supergravity and integrate with respect to all 
variables except the radial coordinate $u$. One is thus lead to an effective classical mechanical system. 
The simplest equation is for the nonextremality function $x$:
\be
\eqlabel{eq:x}
x''=0 \ , \ \ \ \  x= a u \ , \ \ \ \     a=\const .
\ee
The reason for such a simple equation is that, as explained in \cite{105}, 
it does not appear in the effective one dimensional Lagrangian, except for its kinetic term.

The other functions $y,w,z,f$ and $\P$ are to be determined from a coupled system of equations: 
\bea
\label{eq:sys}
10y'' - 8 e^{8y} (6 e^{-2w} - e^{-12 w})   + \P''
&=&0, \nonumber \\ 
10w'' - 12 e^{8y} ( e^{-2w} - e^{-12 w})   - \P'' &=&0, \nonumber \\  
\P''    + e^{-\P + 4z - 4y-4w} (f'^2 -  e^{2 \P + 8 y+8w} P^2)&=&0, \nonumber\\
4z'' -  (Q+ 2 P f)^2  e^{8z} - e^{-\P + 4z - 4y-4w} ( f'^2 +  e^{2 \P + 8 y+8w} P^2) &=&0, \nonumber \\  
(e^{-\P + 4z - 4y-4w} f')' - P (Q+ 2 P f) e^{8z} &=&0.
\eea
The integration constants are 
subject  to the  zero-energy constrain $T+V=0$, i.e.
\bea
\label{eq:constrain}
&&5  y'^2   - 2 z'^2  - 5 w'^2 - {\frac18} \P'^2 
- {\frac14}  e^{-\P +  4z -4y - 4 w } f'^2 
 -   \  e^{8y} ( 6 e^{-2w} - e^{-12 w} ) \nonumber\\
&& +  {\frac14} e^{\P+  4z + 4y + 4 w } P^2 + 
 {\frac18}  e^{8z} (Q + 2 P f)^2   - 3 a^2 = 0 \ . 
\eea
As mentioned while introducing the Ansatz, the function $y$ amounts to a choice of the radial coordinate. This is 
relevant for understanding the dimensions of all quantities. Note that in particular, the system (\ref{eq:sys}) and the 
constrain (\ref{eq:constrain}) are invariant under 
$e^y\to L_0 e^y$ and $u\to L_0^{-4}u$ if we assume that $Q\to L_0^4Q, P\to L_0^2P, a\to L_0^4a$ and $f\to L_0^2f$. 
We therefore express all dimensionfull quantities in units where $L_0=1$. 

\subsection{Some analytic solutions}
In this section we review some known analytic solutions to the system. 
Our goal is to develop the necessary intuition into the properties of the solution we are seeking. The existence 
of exact analytical solutions provides us with the unique opportunity to test the numerical method. 
The numerical treatment of these solutions will be presented in section \ref{sec:numerics} and this section 
can be considered as preparatory.

\subsubsection{The blown up conifold}
One of the simplest class of solutions to the above system has the 
form of $\mathbb{R}^{3,1}\times \mathbb{CY}$, where $\mathbb{CY}$ 
stands for a Calabi-Yau space and more concretely a Ricci-flat 
space. The simplest solution corresponds to the 
conifold and is defined by 
\be e^{4y}=\frac{1}{4u}, \ee
and all other functions are zero or constant in the case of the dilaton. In this 
class of solutions we also find the blown up conifold\footnote{An adjustment in the 
periodicity of $\psi$ is required.} \cite{pt}. 
This solution is usually written as
\be ds_6^2=\kappa^{-1}dr^2 + r^2 \left(\kappa e_\psi^2 + 
e_{\theta_1}^2+e_{\phi_1}^2+e_{\theta_2}^2+e_{\phi_2}^2\right), \ee
where
\be \kappa=1-\frac{b^6}{r^6}=e^{-10 w}, \quad r=e^{y+w}, \quad 
b\le r< \infty. \ee
This space has two nontrivial functions: $y$ and $w$. The direct relation between 
$r$ and $u$ is given by: 
\be
du=-\frac{dr}{r^5(1-\frac{b^6}{r^6})}.
\ee
In general the relationship between the coordinates $u$ and $r$ is nontrivial but for large $r$ we 
recover small $u\sim 1/r^4$.
The meaning of 
$w$ is clearly to squash the $U(1)$ fiber with respect to the 
four-dimensional base.  Note that as $r\to \infty$ the squashing 
vanishes $w\to 0$.

\subsubsection{Non-Extremal  D3-brane solution: Singular and Regular}\label{sssec:nonextremald3}

The general form of the nonextremal solutions contains a {\it 
singular} horizon. We should consider it here to show how the 
singularity looks in the numerical analysis. Imposing regularity of 
the horizon of the solution leads us to the {\it standard} 
nonextremal D3 brane 
(henceforth we will call it just the `D3 solution'). 

Let us  consider   
the general system of equations (\ref{eq:sys})
with $P=0$ and $f=0$, that is, with no varying flux and also with $\P=0$
and $w=0$.
Then,  we are left with (\ref{eq:x}) and 
  the following  system:

\be
 y''  - 4 e^{8y} =0 \ , \ \ \ \ \
z''   - {\frac14} Q^2 e^{8z} =0  \ , 
\ee
i.e.  
\be
 x'=a \ , \ \ \ \ \ 
y'^2 = b^2  + e^{8y} \ , \ \ \ \ \ 
z'^2 = c^2   + q^2 e^{8z} \ , \ \ \ \  \ \  q\equiv  {\frac14} Q \ , 
\ee
with the  integration constants  $a,b,c$ 
related  by the zero-energy constraint 
\be
\eqlabel{constraind3}
5b^2 - 3 a^2 - 2 c^2 =0   \ . 
\ee

Assuming that $a,b,c\geq 0$ to preserve asymptotic conditions, the general solution is:  
\be
e^{4y} =  \frac {b}{\sinh 4b u } \ , \ \ \ \ \ 
 e^{4z} =  \frac {c}{q\  \sinh 4 c (u + k)   } \ ,\ \ \ \ \ \ 
e^{4x} = e^{4au} \ , \ \ \ 
\label{eq:neD3}
\ee
where $k$ is defined by $
e^{ 4c k} = q^{-1}
 ( \sqrt{ q^2 + c^2} + c ) \equiv \g \ $.  
Note that the integration constant $k$ implies just a shift of variable $u$.
To meet AdS asymptotic conditions we must choose $k=0$. 

In terms of the familiar D3 brane Ansatz (\ref{eq:d3}):
\be
\r^4= e^{4y+4x} =  \frac{ 2b e^{4(a-b)u}} {1 - e^{-8bu} } \ , \ \ \ 
\ \ \ \ \    g_1=g_2=g= e^{-8au} \ , 
\ee
\be h= e^{-4z- 4x} =  \frac{q}{2c}
e^{4(c-a)u}( 1 - e^{- 8c u} ) 
    \ . 
\ee
At small $u$ (large $\r$) we have 
\be
g= 1 - \frac{2a}{\r^4} + ... \ ,  \ \ \ \ 
h = \frac{q}{ \r^4} + ...  
\ , \ \ \ \    \r^4 = \frac1{4 u} + ... \ . 
\ee

The general solution with arbitrary $b$ and $c$ 
reduces to the standard extremal D3-brane background
only if we set $b$ and $c$ proportional to $a$,
satisfying the constrain (\ref{eq:constrain}).
As we have emphasized several times, for arbitrary values of $b$ and $c$ 
the solution is singular.
To see why that is the case,
let us consider a singular case, discussed at length in \cite{105}, to better 
understand the role of parameters in the singularity property of the solution. 
The simplest  special case  is $c=0$   where $z$ 
satisfies the 1-st order equation $z' = - q e^{4z}$.
One can solve the system (\ref{eq:sys}) to get 
\be
e^{4y} =  \frac{b}{\sinh 4b u  } \ , \ \ \ \ \
 e^{-4z} = 4 q u   \ ,\ \ \ \ \ \
e^{4x} = e^{4au} \ ,  \ \ \ \  \ \ b=\sqrt \frac35  a\ , 
\ee
and thus 
\be
\r^4= e^{4y+4x} =  \frac{ 2b e^{4(a-b)u}}{ 1 - e^{-8bu} } \ , \ \ \ \  
 h= e^{-4z- 4x} =  4 q  u  e^{-4au}
 \ , 
\ \ \  \   g= e^{-8au} \ . 
\ee
Note that $\rho$ is not well-defined as a radial coordinate for $b\ne a$. 
This solution
has a  singular horizon at $u=\infty$ as described in \cite{105}. Using the expression 
for $R_{mnkl} R^{mnkl}$  one concludes that the metric is singular. 
Most importantly,
 the area of the horizon, 
 which is proportional to $\exp(-2z+3x+5y)$  vanishes as $u\to \infty$. 
Namely, it is proportional to $u^{1/2}\exp(-(\sqrt{15}-3)u) \longrightarrow 0$ as $u\to \infty$. 
{\it One of the main criterium for the solution we are seeking is thus,  to have a 
finite nonzero area of the horizon.} We have seen that the choice of the constants $a, b$ and $c$ plays an 
important role in fulfilling this condition.  

For small $u$  (large  distances) 
and in the limit $a\to 0$, we still get the
standard asymptotic extremal D3-brane behavior

\be
\r^4 = \frac1{ 4 u} + ... \ ,  \ \ \ \ \ 
g= 1 - \frac{ 2a }{ \r^4} + ... \ , \ \ \  
h = \frac{ q}{ \r^4} + ...  
\ .
\ee
Note that in the large$-\rho$ asymptotic the solution is like the standard D3.
Most of the problems, as was pointed out, are localized at the horizon.

The  standard  non-extremal  D3-brane solution 
corresponds to the  case  when 
\be
b=c=a  \  , 
\ee
This condition, together with the constrain (\ref{constraind3}) means that {\it regularity picks a line in the  
two dimensional surface which is the space of solutions}. We will see that something similar happens in the solution we 
construct in section \ref{sec:new}. The solution of the system (\ref{eq:sys}) takes the form: 
\be
\eqlabel{eq:regd3}
e^{4y} =  \frac{ a}{ \sinh 4a u  } \ , \ \ \ \ \ \
 e^{4z} =  \frac{ a}{  q \sinh 4 a u  } \ ,\ \ \ \ \ \ \
e^{4x} = e^{4au} \ ,  
\ee
Note that  near the horizon ($u\to \infty$) 
\be
y= y_* - a u +  \frac 14  e^{-8au} + O( e^{-16au}) \ , \ \ \ \ \
z= z_*  - a u + \frac 14  e^{-8au} + O( e^{-16au}) \ ,
\ee
\be
 y_* = \frac 14 \ln 2 a \ , \ \ \ \ 
z_* = \frac 14 \ln \frac{2 a}{ q} \ , \ \ \
\ee
We can verified that this solution has finite area of the horizon. 
Indeed $\exp(-2z+3x+5y) \to \exp(-2z_*+5y_*)$ which is finite, 
$q^{1/2}(2a)^{3/4}$,
as we approach the horizon $(u\to \infty)$. 
We see that the requirements on the asymptotics 
of the functions $y$ and $z$ in order to have a horizon with finite  area are very stringent.

A more recognizable form of this solution is given as 
\be
ds^2 =  h^{-1/2} ( g dX_0^2 + dX_i dX_i)
+  h^{1/2}  [ g^{-1}  d\r^2 
+ \r^2 (d M_5)^2] \ , \ \ \ \ 
\ee
\be
g= e^{-8x} = 1 - \frac{2 a}{  \r^{4}}  \  , 
 \ \ \ \ \ \ \ 
  \r^4= \frac{ 2 a }{1- e^{-8a u} } \ ,
\ \ \ \ \  h=  e^{-4z- 4x}= \frac{ q }{  \r^{4} }  \ . 
\ee

Let us clarify the relationship between the radial coordinate $u$ and the more 
standard coordinate $\rho$. Using $2a=\rho_0^4$,  we have
\be
\eqlabel{eq:urho}
du =\frac{d\rho}{\rho^5}\left(1-\frac{\rho_0^4}{\rho^4}\right)^{-1}.
\ee
This can be integrated to 
\be
\eqlabel{eq:urhohor}
u=-\frac{1}{\rho_0^4}\ln \left(1-\frac{\rho_0^4}{\rho^4}\right).
\ee
Note that in the domain of $\rho_0 \le \rho < \infty$ we have that $u$ ranges in $0 <u<\infty$. 
The position of the horizon which is finite in the $\rho$ coordinate becomes infinite in 
the $u$ coordinates. 

\subsubsection*{Some thermodynamics and universality of regular nonextremal D3 brane horizons}
For the standard nonextremal D3 brane the horizon is located at $\rho=\rho_0$, and the area is given by:
\be
A=V \omega_5 R^2 \rho_0^3, 
\ee
where $V$ is the volume due to the coordinates $X^i$, $R$ is the radius 
of AdS and appears in the harmonic function as $h=R^4/\rho^4$, $\omega_5$ is the volume of $T^{1,1}$. 

The natural temperature associated with the nonextremal D3 is obtained from the regularity of the 
Euclidean section:
\be
T=\frac{\r_0}{\pi R^2}.
\ee
Of course, the local temperature is $T_{local}=T R\rho/\sqrt{\rho^4-\rho_0^4}$ and decreases as
$\rho\to \infty$. More explicitly, the nonextremal D3 brane metric is of the form
\be
ds^2=\frac{\r^2}{R^2}\left((1-\frac{\r_0^4}{\r^4}) d\tau^2 +dX_i dX^i\right)
+\frac{R^2}{\r^2}\left(1-\frac{\r_0^4}{\r^4}\right)^{-1}d\r^2 +R^2 d\o_5^2.
\ee
Near the horizon one can introduce the following coordinate
\be
\rho=\rho_0(1+\frac{r^2}{R^2}).
\ee
In terms of this radial coordinate in the limit of $r\to 0$ the relevant part of the metric takes the form
\be
ds^2=dr^2+r^2\left(\frac{2\rho_0}{R^2}d\tau\right)^2.
\ee
For the angular part to have period $2\pi$ we obtain the above temperature as the inverse of the 
period: $T=1/\beta$.

In the $u$-coordinates the area of surface defined by a horizon at $u={\rm constant}$ is
\be
A=V\o_5 \exp\left(-2z+3x+5y\right).
\label{eq:horA}
\ee
Given that the equation of motion for $x$ has the general solution $x=a u$ we are forced into the following situation. 
If the horizon is at $u\to \infty$, then in order for the area $A$ to be finite we need the following asymptotics for 
$z$ and $y$:
\be
z\to \alpha\, a u + z_*, \qquad y\to \beta\, a u + y_*, 
\label{eq:yzasym}
\ee
with the condition that 
\begin{equation}
-2\alpha+5\beta=-3
\label{eq:yzcond}
\end{equation}
Note that the regular nonextremal D3 brane corresponds to $\a=\b=-1$. 
The main claim is that: {\it The existence of a regular horizon fixes the asymptotic behavior of the metric coordinates
$x,y$ and $z$.}

Similarly, one can obtain an expression for the temperature. Namely,  the relevant part of the metric is 
\be
ds^2=e^{2z-6x}d\tau^2+e^{-2z+10y}du^2.
\ee
We introduce a new radial coordinate as:
\be
\r=e^{z-3x}.
\ee
We can now rewrite the metric as:
\be
ds^2=\frac{e^{-4z+10y+6x}}{(z'-3x')^2}\bigg[d\r^2+\r^2\left(e^{2z-5y-3x}(z'-3a)\,\, d\tau\right)^2\bigg].
\ee
Note that, again, {\it the requirement of finite temperature fixes the large-$u$ asymptotic of various metric functions to 
be $2z-5y-3x \to {\rm constant.}$} Imposing absence of conical singularity
 we find that the temperature defined as the inverse of the period is
\be
T=\frac{|\a-3a|}{2\pi}e^{2z_*-5y_*},
\label{eq:BHT}
\ee
where we used that near the horizon the asymptotic form of $z \sim \a a u + z_*$ and similarly for $y$.

\subsubsection{The Klebanov-Tseytlin background}
An important property of this solution is that it describes the asymptotic of most known 
solutions with $P\ne 0$ and  for small values of $u$, see for example \cite{kn,kt,ks,resolved,pt,papa,butti}. 
The  KT solution  can be represented as:
 
 $$  x=0\ , \ \ \ \ \ \ \ w=0\ , \ \ \ \ \P=0 \ , $$ 

\be  e^{-4y} = 4u \ , \ \  \ \ \ \ \ \ 
\ f= f_0  - \frac{P}{4} \ln u \ ,   
\ee
$$ e^{-4z} =  K_0 u  
 -  \frac{P^2}{2}  u ( \ln u -1) \ ,  \ \ \ \ \ \ \   
 K_0 = Q + 2 P f_0 \  , $$ 
i.e. 
\be
\eqlabel{eq:zkt}
e^{-4z} = h =  (Q + 2 P f_0 + \frac{P^2}{2} ) u  
 -  \frac{P^2}{2}  u \ln u \ .     
\ee
For convenience, in the case of $P\ne 0$ we can parameterize the function $K$ as
\be
K(u)=-\frac{P^2}{2}\ln (u L_P^{-4}).
\ee

As mentioned before the KT solution is an attractor of sorts. Let us clarify to what extend it 
is generic. The main claim is the following:
{\it Any solution of the class we considered with $P\ne 0$ will asymptote to the 
KT solution if we impose a constant value of the dilaton in the asymptotic region.}
We will not proceed to give a formal proof\footnote{A rigorous analysis of fixed points of the 
system (\ref{eq:sys}) will be presented elsewhere.} of the above statement, instead we will show how it comes about. 
If we impose the conditions of $\Phi=0$ and $w=0$ in the system of equations we find that
the equation for $y$  yields:
\be
y''=4e^{8y},
\ee
while the equation for the dilaton in (\ref{eq:sys}) implies that
\be
f'^2=P^2 e^{8y}.
\ee
One can explicitly solve 
this system and obtain that as $u\to 0$ one has $f\sim \ln u$. 

Let us present the ten-dimensional analysis which provides an interesting perspective. 
For very large values of $\rho$ we 
have that the metric becomes essentially a cone over $T^{1,1}$, in the effective one dimensional system language 
this is equivalent to having $w=0$. The metric is thus
\be
d\rho^2+\rho^2 ds^2(M_5).
\ee
Imposing that we have $P\ne 0$ means that we are introducing a  flux described by 
\be
F_3=P \Omega_3.
\ee
If we also assume that the dilaton is constant we have and equation between the forms $F_3$ and $H_3$ which can schematically be
written as $|F_3|^2=|H_3|^2$. This equation can also be understood as a consequence of having an imaginary self-dual 
$G_3=H_3+iF_3$.  Basically, it can be understood as a consequence of $*_6 \; F_3=H_3$, where subscript means that
we consider the Hodge dual in the six dimensional space. This is a very simple equation to 
analyze. In fact, one has
\be
*_6 \, F_3=*_6 \, P \Omega_3=P \frac{d\r}{\r} \Omega_2.
\ee
Thus, we determine that $B_2=P\,\Omega_2 \ln \r$,  where $\Omega_2$ is some 2-form Hodge dual to $\Omega_3$. This logarithm in $B_2$ propagates to a logarithm in the metric via the 
equation for $F_5$ which is solved by
\be
K(\rho)=Q+2P^2 \ln \r/\r_0.
\ee
Clearly, this solution has a repulsive singularity around $\rho_{sing}=\rho_0\, e^{-Q/2P^2}$ which is locate at small values of 
$\rho$ meaning large values of $u$ (see appendix \ref{geodesics} for more details on the causal structure).

\subsubsection{A singular generalization  of the KT Solution}
This solution is obtained by assuming that the dilaton is constant and $w\equiv 0$ in (\ref{eq:sys}) and allowing 
$a\ne 0$. The solution for $y$ takes the form of 
\be
e^{4y} = \frac{ b}{ \sinh 4bu} \ ,
\ee
with $b=a\sqrt{3/5}$. Knowing $y$ allows us to determine the other variables and one finds:
\bea
f&=& f_*  -  \frac{P}{ 4} \ln \tanh 2bu \ ,   \nonumber \\
e^{-4z} &=&K_* u + \frac{ P^2}{8 b} \bigg({\rm Li}_2(-e^{-4bu})
- {\rm Li}_2(e^{-4bu}) \bigg) \ , 
\eea
and we recall that $K(u)=Q+2Pf(u)$. This solution was originally obtained in \cite{alex} as a nonextremal generalization 
of the KT solution. 
It was subsequently studied in \cite{105}, where it was established that it has a horizon which coincides with a 
singularity at 
$u=\infty$ and that it reduces to a singular nonextremal D3 brane for $P\to 0$.

\subsubsection{Perturbative solution with varying flux around nonextremal D3}
In this section we review a solution presented in \cite{172}. This solution is obtained as a 
first order correction in $P$ to the nonextremal D3 brane background.  
The main idea is to perform a perturbation around the regular nonextremal D3 brane $(P=0)$, in the regime where
$P^2/K_* \ll 1$ and it serves as a small parameter. 

It is therefore convenient to rescale  the
fields  by 
appropriate powers of $P^2$, setting  
\be
K(u)= K_* + 2 P^2 F(u)
 \ , \ \ \ \ \ \P (u) = P^2 \p(u) \ , \ \  \ \ w(u) = P^2 \omega (u) \ ,
\ee
and 
\be
y\to   y  +  P^2 \xi\ , \ \ \ \ \  
e^{-4z} \to  e^{-4z}  + P^2  \zeta \ ,  \ \ \  
{\rm i.e.} \ \ \
 z\to   z  +  P^2 \eta\ ,  \ \ \ \  \zeta
= -4 e^{ - 4 z} \eta  + O(P^2) \ ,  
\ee
where  $y, z$  represent 
 the pure D3-brane solution (\ref{sssec:nonextremald3})
$ e^{-4y} = a^{-1} \sinh 4au,\ \ e^{-4z} =
 \frac{K_*}{4a}\sinh 4a u$,  and $\xi$ and $\zeta$
or $\eta$
are  corrections to it.
The system (\ref{eq:sys})
takes the following explicit form:
\bea
10\xi'' - 320  e^{8y} \xi 
   +  \p''  +O(P^2)    &=&0 \ , \nonumber\\
10\omega''  -  120 e^{8y} \omega - \p'' +   O(P^2)  &=&0
\ , \nonumber \\
\p'' + e^{ 4z - 4y} (F'^2 - e^{  8 y}) + O(P^2)&=&0 \ , \nonumber \\
(e^{ 4z - 4y} F')' -  K_* e^{8z}  + 
 O(P^2)&=&0 \, \nonumber \\
4\eta'' - 8 K_*^2  e^{8z}\eta 
 -  4  K_*  F e^{8z}
 -    e^{ 4z - 4y} ( F'^2 +  e^{ 8 y})
 +O(P^2) &=&0 
\eea
The constraint \ref{eq:constrain} becomes 
\be
10 y' \xi'  - 4 z' \eta' 
- \frac 14   e^{ 4z - 4y}  F'^2
- 40 e^{8 y} \xi 
+ \frac 14   e^{ 4z + 4y}
+    K_*^2 e^{8 z}  \eta   + \frac 12 K_* e^{8 z} F
+ O(P^2) = 0 \ . 
\label{eq:pertconstrain}
\ee
The solution takes the form
\bea
\label{kugeneral}
K(u)&=&K_*-\frac{P^2}{2}\ln\left(1-e^{-8au}\right), \\
\p&=&\p_*+\frac{1}{4K_*}{\rm Li}_2(e^{-8au}), \qquad \p_*=-\frac{\pi^2}{24K_*}. \nonumber 
\eea
The rest of the fields can be expressed in terms of a new radial coordinate of the form 
\be
v=1-e^{-8au}.
\ee
In this case we obtain:
\bea
\xi&=&\frac{1}{20K_*}+ \frac{1}{40K_*\, v}\left((-2v+(v-2)\ln(1-v))\ln v + (v-2){\rm Li}_2(v)\right), \nonumber \\
\eta&=&\frac{v-2}{16K_*\, v}\left(\ln v \ln (1-v)+{\rm Li}_2(v)\right).
\eea
Similarly, one obtains a simple equation for $\omega$ (see \cite{172} for details).

\subsubsection*{Regime of validity, extrapolation and thermodynamics}

The above solution was constructed with the assumption that  $P^2/K_*\ll 1$. Since we 
considered only the linearized approximation there is an intrinsic limit on the values of $u$. Recall 
that for nonextremal D3 branes $K(u)=Constant\sim K_*$ which counts the number of D3 branes. For large values of 
$u$, the solution (\ref{kugeneral}) becomes 
\be
K(u) \sim K_* +\frac{P^2}{2}e^{-8au}.
\ee
The second term is subleading for  large $u$ and thus in this region the solution remains valid. As we 
decrease the value of $u$, we reach a point where the first and second term in $K(u)$ given by (\ref{kugeneral}) 
are of the same order. This 
happens for 
\be
u_{c}=-\frac{1}{8a}\ln (1-e^{-2K_*}{P^2})\approx \frac{1}{8a}e^{-2K_*/P^2}.
\ee
The value $u_c$ is small but nonzero.
Altogether, the {\it regime of validity } is: 
\be
\eqlabel{eq:regime}
P^2/K_*\ll 1, \qquad {\rm and} \quad u\gg u_{c}=\frac{1}{8a} e^{-2K_*/P^2}.
\ee

Nevertheless, the authors of \cite{172} decided to explore the small $u$ regime and found that 
\be
\label{eq:kuzero}
K(u) \sim K_*-\frac{P^2}{2}\ln (8au).
\ee
Amazingly, this form of the solution resembles  the KT solution. 
The main idea of \cite{172} is to match the $u\to 0$ asymptotics of the solution (\ref{eq:kuzero}) with 
the KT solution (see equation 3.8 of \cite{172})
\be
\label{eq:kukt}
K_{KT}=-\frac{P^2}{2}\ln (uL_P^4)
\ee
Matching these two solutions, that is (\ref{eq:kuzero}) and (\ref{eq:kukt}) we find that (5.15) of \cite{172})
\be
\label{eq:ak*}
8aL_p^{-4}=e^{2K_*/P^2}.
\ee
This is the relationship between $K_*$ and $a$. Finally one has
 the relation between $a$ -- the nonextremality parameter and 
the temperature\footnote{This is an approximate relation that ignores some issues of asymptotics.}. Formula 
(5.16) in \cite{172} is nothing but the temperature of nonextremal D3 branes as a function 
of the radius of the horizon $a^{1/4}$ and the number of D3 branes $K_*$. Thus, we quote
\be
T\sim a^{1/4} K_*^{-1/2}.
\ee
Plugging this relation into (\ref{eq:ak*}) we find that to leading approximation
\be
K_*\sim \frac{P^2}{2}\ln \frac{T}{\Lambda},
\ee
which is quoted in (5.44) of \cite{172}. This expression is crucial in understanding the thermodynamics of 
this class of solutions. One basically conjectures that
the entropy per unit volume satisfies 
\be
\frac{S}{VT^3} \sim \frac{P^4}{L_P^8} \left(\ln \frac{T}{\Lambda}\right)^2.
\ee
A similar formula was recently discussed in the context of holographic renormalization in \cite{aby}.

\subsubsection*{Not all is well with extrapolating the perturbed nonextremal D3 brane  solution}\label{sss:breaking}
This perturbative analysis is a very valuable tool to understanding the thermodynamics. However, it has
several shortcomings. First, the thermodynamics requires the understanding of the region where the 
perturbation breaks down. Second, one can confirm that there are obstructions to extending the solutions 
analytically past its regime of validity. One can check that the constrain is not satisfied. In fact, expanding 
the constraint (\ref{eq:pertconstrain}) for small u gives a divergent term already at first order in $P^2/K_*$:
\be
-\frac{a}{2u}-\frac{1}{3}a^2+\frac83 a^3 u +{\cal O}(u^2).
\ee

\section{Numerical analysis of known solutions}\label{sec:numerics}

To search for numerical solutions we rewrite  (\ref{eq:sys})  as a system of first order differential equations. 
To ensure that the numerical solutions automatically satisfy the Hamiltonian constrain, the corresponding equations were modified to include explicitly the information given by expression (\ref{eq:constrain}). 
The resulting system contains ten coupled fields 
\{${y}(u), {z}(u), {w}(u), {\Phi}(u), {f}(u), \tilde{y}(u), \tilde{z}(u), \tilde{w}(u), \tilde{\Phi}(u), \tilde{f}(u)$\} 
related by the corresponding non-linear first order differential equations:
\begin{eqnarray}
\frac {dy}{du} &=& \tilde{y} ,\hspace{1cm}
{\frac {dz}{du}}  =\tilde{z} ,\hspace{1cm}
{\frac {dw}{du}}  =\tilde{w} ,\hspace{1cm}
\frac {d\Phi}{du} = \tilde{\Phi} ,\hspace{1cm}
\frac {df}{du} = \tilde{f} ,\\
\frac {d\tilde{y}}{du}   &=& 2\tilde{y}^{2} -\frac 45\,\tilde{z}^{2}-2\, \tilde{w}^{2}
-\frac1{20}\,\tilde{\Phi}^{2} \nonumber\\
&+& \frac25\,{e^{8y}} \left( 6{e^{-2w}}-{e^{-12w}} \right) +\frac1{20}\, \left( Q+2Pf\right)^{2}{e^{8z}}-\frac65{a}^{2},\\
\frac {d\tilde{z}}{du} &=&5\tilde{y}^{2}-2\tilde{z}^{2}-5\tilde{w}^{2}-\frac18\tilde{\Phi}^{2}\nonumber\\
&+&\frac38 \left( Q+2Pf \right)^{2}{e^{8z }}
-{e^{8y}} \left( 6{e^{-2w}}-{e^{-12w}} \right) +\frac12{e^{\Phi +4z +4y + 4w}}{P}^{2}-3{a}^{2},\\
\frac {d\tilde{w}}{du} &=& -2\tilde{y}^{2}+\frac45 \tilde{z}^{2}+2\tilde{w}^{2}+\frac1{20}\tilde{\Phi}^{2}\nonumber\\
&+&\frac85{e^{8y}} \left({e^{-2w}}-{e^{-12w}} \right)
-\frac1{20}\left( Q+2Pf \right)^{2}{e^{8z}}+\frac65{a}^{2},\\
\frac {d\tilde{\Phi}}{du}  &=& -20 \tilde{y}^{2} + 8 \tilde{z}^{2} + 20 \tilde{w}^{2} + \frac12 \tilde{\Phi}^{2}\nonumber\\
&+& 4{e^{8y}} \left( 6{e^{-2w}}-{e^{-12w }} \right) - \frac12 \left( Q+2Pf \right) ^{2}{e^{8z }}+ 12{a}^{2},\\
\frac {d\tilde{f}}{du}   &=& \left( \tilde{\Phi} -4\tilde{z} +4\tilde{y} +4\tilde{w}  \right)\tilde{f} 
+P \left( Q+2Pf   \right) {e^{\Phi +4z  +4y  +4w  }},
\end{eqnarray}
where 
$P$, $Q$ and $a$ are the three parameters that determine the behavior of the solutions.
To solve it we use a combination of the methods implemented in the Maple software
to find numerical solutions of ordinary differential equations\footnote{
Our Maple sheets are available upon request.}. 
For most calculations we have used the seventh-eight order continuous Runge-Kutta method 
which, thanks to its adaptive scheme, 
provides a great control upon the output accuracy. 
In those cases where the stiffness typical of singularities was present, 
we switched to the Livermore Stiff Ode solver. 
This also allowed us to get some information about the stability of the obtained solutions
with respect to the truncation error.

The option of yielding the output of the numerical computation as a list of procedures
is, in general, not very economical for long and cyclical calculations. 
However, it 
proved to be convenient for the study of those quantities depending on the metric functions and on the dilaton such as 
those describing the thermodynamics.

To use a numerical solver, one first needs to set up its arguments 
with values providing outputs as accurate as required.
We recall that in an adaptive scheme 
the discretization of the independent variable domain is automatically refined 
until a theoretical measure of the error 
goes below some tolerance previously fixed by the user.
So, to set the optimal tolerance 
we compared the known analytical solutions presented
in the previous section with numerical outputs, 
using the measure given by,
\begin{equation}
\chi=\sqrt{\sum \left[v^{an}_i(u)-v^{num}_i(u)\right]^2 }\, , \quad i=1\cdots10\, .
\label{eq:xi2}
\end{equation}
Here $v^{an}_i(u)$ and $v^{num}_i(u)$ stand 
for the value at $u$ of each of our ten variables 
as given, respectively, by the analytical and numerical solutions.
Next, we must determine when this measure can be taken as negligible.
We say that $\chi$ is negligible if adding its value to one of the parameters
entering our system does not lead to a qualitative departure from the analytical solution
corrresponding to the unchanged value of the given parameter.
For instance, starting with a very low tolerance and $a=0$ 
we gradually incremented $a$ until we found a qualitative departure from the KT solution.
Then, we increased the tolerance 
looking for the maximal value that keeps the just described situation essentially unchanged.
The same procedure was carried out with $P$ and the standard non-extremal D3 solution.
This way 
we found that for a tolerance of $10^{-14}$, 
$\chi_{max}=10^{-10}$ can be safely regarded to be negligible.

\begin{figure} 
\centerline{\includegraphics[width=0.55\linewidth]{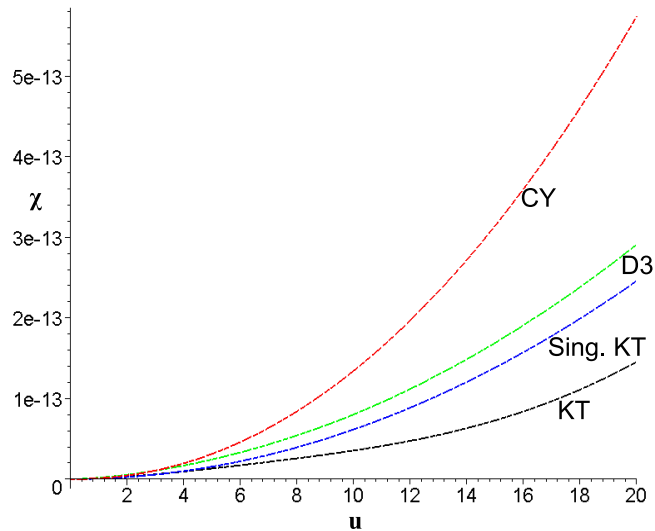}} 
\caption{Plot of the difference between analytical solutions and the corresponding numerical outputs:
$\mathbb{R}^{3,1}\times \mathbb{CY}$ (red), 
Klebanov-Tseytlin (black), standard nonextremal D3 (green) and 
singular nonextremal generalization of Klebanov-Tseytlin (blue).}  
\label{fig:fig1} 
\end{figure}

In figure \ref{fig:fig1} 
typical errors are presented for the following solutions:
KT (black), singular nonextremal generalization of KT (blue), 
D3 (green)  and $\mathbb{R}^{3,1}\times \mathbb{CY}$ (red).
The error measure is given by Eq.(\ref{eq:xi2}) and the plots
show that, after tuning up the numerical method, 
the calculations typically yield reliable results for a wide range of values of the independent 
variable $u$ in the interval $(10^{-3}, 20)$, and of the involved parameters. 

\begin{figure} 
\centerline{\includegraphics[width=0.55\linewidth]{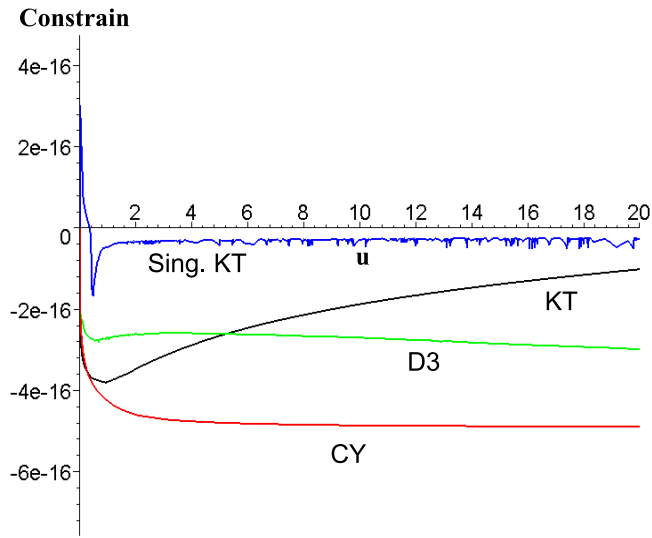}} 
\caption{Plot of the numerical output of the constrain for four study cases:
$\mathbb{R}^{3,1}\times \mathbb{CY}$ (red), 
Klebanov-Tseytlin (black), standard nonextremal D3 (green) and 
singular nonextremal generalization of Klebanov-Tseytlin (blue).}  
\label{fig:fig2} 
\end{figure}

In the absence of a pattern solution to check for the accuracy of the computations,
we fortunately have the Hamiltonian constrain. 
Since our system was explicitly constructed to yield only solutions satisfying equation ({\ref{eq:constrain}}),
the necessary fulfillment of this condition could be readily tested for every case we solved.
As an example, we present in figure \ref{fig:fig2} 
the numerical results for the constrain corresponding to the four study cases mentioned above.

\subsection{Understanding the numerical solution}\label{ssec:unumbers}
Next, we move into understanding the output of the numerical computation. Our goal is to establish the 
existence of solutions with a nonsingular horizon and therefore we need to develop the appropriate criteria at $u\to \infty$. 
At the same time we aim to make a definite statement about the behavior of the solutions for small values of $u$.

It is important to note that since this is a nonlinear system of differential equations,
the asymptotic form of the solutions is typically very sensitive to the values of the ten `initial' conditions.
So, to search for a solution with given behavior at small and at large values of $u$
by just guessing the conditions at any of these boundaries it is an unfruitful task.
The simplest way of being certain of the existence of a black hole with any $u\to 0$ asymptotics
is the following.

First, let us recall that in $u$-coordinates the horizon is located at infinity.
Therefore, we face the difficulty of estimating how representative some given numbers are
of the actual solution near the horizon.
Fortunately,
assuming that the nature of the horizon is of the type of nonextremal D3-brane we can use formula 
(\ref{eq:urhohor}) to 
estimate how close to the horizon we are. Numerically, we will consider a sequence for which $\rho$ approaches the 
horizon as $\rho=\eta \rho_0$. Assuming that $\rho_0^4= 2a$ we find that
\be
2 u_{\eta} a = -\ln (1- \eta^{-4}).
\label{eq:u90}
\ee 
This quantity characterizes how close we are to the horizon. 
Note that 
\be
\frac{\rho-\rho_0}{\rho_0}=\eta-1.
\ee
Alternatively, we can look at this formula as giving a prescription that tells us for different values of $a$ how 
far we have to go in $u$ to reach the same ``distance'' from the horizon, where we refer to distance 
as a concept on $\rho$. 
Hence, we use expression (\ref{eq:u90}) to estimate the value of $u$ corresponding, for instance,
to the $90\%$ of the distance to the horizon.

Next, we set the initial conditions for our variables at this value of $u_{90\%}$,
using expressions (\ref{eq:neD3}) for the solution of the non-extremal D3 case.
We then integrate the system forward and identify numerically the presence of a horizon. 
To do that we developed the following set of criteria.

Let us start by considering a massive geodesic in the general metric (\ref{eq:metric}). 
The effective Lagrangian for radial motion is given by 
\be
\call=-e^{2z-6x}\dot{t}^2 + e^{-2z+10y}\dot{u}^2.
\ee
Normalizing the effective Lagrangian to be $\call =-1$ and using the fact that the Lagrangian 
does not explicitly depends on time we find an equation for the radial coordinate:
\be
\dot{u}^2 +e^{2z-10y}=E^2 \, e^{6x-10y}.
\ee
This equation does not readily allows an interpretation in terms of classical motion of a particle 
with energy $E$ in a potential. For this aim, we find it useful to define a new variable $v$ given by
$ dv= e^{-3x+5y}du$. 
The geodesic equation then becomes
\be
\label{eq:VEff}
\dot{v}^2+e^{2z-6x}=E^2,
\ee
where the functions $x,y$ and $z$ are now viewed as functions of $v$.  
In this presentation we can consider 
the function $e^{2z-6x}$ as an effective potential describing the classical motion of a particle. 
Using this analogy
and that $dv/du>0$, 
we conclude that if the potential has a wall at some large value of $u$ it indicates a singularity and potential 
repulsive behavior of the form characteristic to the negative mass Schwarzschild or KT. It thus works as a {\it necessary condition} 
for the existence of a black hole that the potential develops no wall for large values of $u$. 
In fact, the case of the regular non extremal D3 brane shows that the potential actually vanishes near the
horizon. We will thus, {\it use the vanishing of the effective potential as a first signal for a possible horizon}. 

Vanishing of the effective potential is not 
sufficient to declare the existence of a horizon at that point. 
We have to take into account the fact that basically $g_{00}\sim e^{2z-6x}$ and therefore
the potential can go to zero and the asymptotic time needs not diverge. The possibility is given by some 
integrable singularity in the expression for the asymptotic time. 

This way, to clarify the existence of a horizon we next consider the proper and asymptotic times. 
For a {\it massive} particle with energy $E$ we obtain that the affine parameter is given by
\be
\tau=\int^{u}_{u_{l}} \frac{e^{5y}}{\sqrt{E^2 e^{6x}-e^{2z}}}du, 
\label{eq:affineT}
\ee
On the other hand, the asymptotic time is 
\be
t=\int^{u}_{u_{l}} \frac{e^{5y}}{\sqrt{E^2 e^{6x}-e^{2z}}} e^{-2z +6x} du. 
\label{eq:asymptT}
\ee
Since we are assuming an AdS configuration, 
in the above definitions there are some subtleties regarding the choice of the lower limit
to ensure that the particle is in the energetically allowed region.
However, this will amount to a numerical correction which will not modify
the convergence of the above integrals when $u\to \infty$.

To have the asymptotic infinity causally disconnected from the region behind the horizon we 
demand the affine time to be finite while the asymptotic time must be unbounded. 
For the numerical calculation of the above integrals  
we decided to implement a version of the
3-5 points adaptive Simpson's rule intended for controlling the accuracy 
while making an optimal use of the
output procedures calling for the solution of the differential system.

Finally, we check for the the finiteness of the horizon area.
Note at this point that, 
according to the analysis of subsection \ref{sssec:nonextremald3},
we have the possibility of doing a cross-check of the accuracy of the numerical output 
as well as of the assumptions made for that analysis.
What we have to do is to observe the asymptotics of variables $y$ and $z$ 
and, if the horizon area and the corresponding temperature are finite, 
then $y$ and $z$ must exhibit a linear behavior when $u\to \infty$,
and, 
for a given $a$, 
the values of the corresponding slopes should obey condition (\ref{eq:yzcond}).

Having found a regular black hole, 
we integrate backward toward $u=0$. 
In order to have an idea of the analytical behavior hidden behind the numbers,
we try several different fits by solving least-squares problems of the difference between
objective functions and the numerical results.
In the case of variables with regular behavior at the origin, 
we find the coefficients of a truncated Taylor expansion.
In those cases with singular behavior, we use a set of singular real functions
which includes the logarithm, the square root and rational functions.
As a cross check, we also looked for the best fit of the corresponding derivatives.

\section{Numerical solutions with varying flux and nonextremal D3-like horizon}\label{sec:new}

Following the recipe in the previous section we were able to find several solutions
with varying flux and nonextremal horizon of the form of D3 branes. 

\subsection{Behavior near the horizon}\label{ssec:BH}

Recall that we are using the nonextremal D3 solution to fix the boundary conditions at $u_{90\%}$.
In other words, first we look for a certain point of the state-space 
(the space spanned by our ten variables) 
crossed by a trajectory
when $P=0$. 
Next, we assume that, 
after setting $P\neq 0$,
close enough to that point there still remain trajectories leading to a black hole solution. 
We guess the position of a new point in one of these trajectories by looking to the corrections
of the original point coordinates
necessary for the Hamiltonian constrain (\ref{eq:constrain}) still being satisfied.
It turned out that this methodology is very effective on the search for regular black holes.

\begin{figure} 
\centerline{\includegraphics[width=0.55\linewidth]{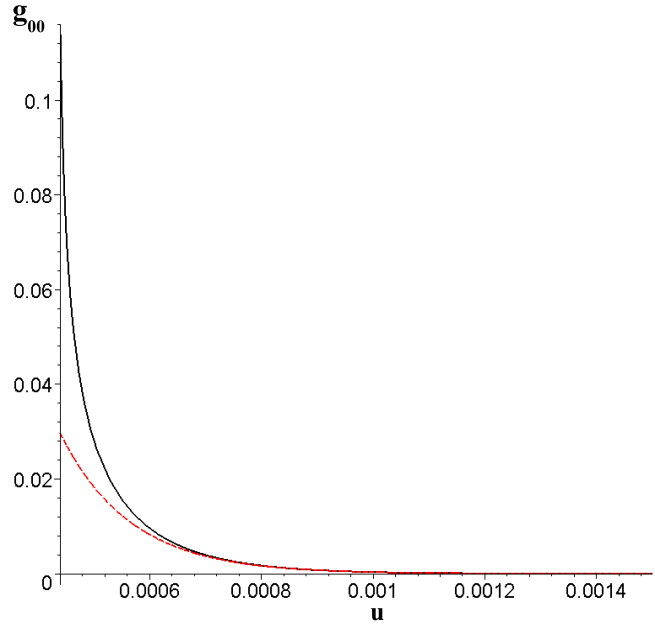}} 
\caption{Plot of the $g_{00}$ component of the metrics (\ref{eq:metric}). 
The solid black curve represents the numerical solution with $P=a=1000$ and $Q=1$,
and the dashed red curve the analytical solution with $P=0$, $a=1000$ and $Q=1$.
Here $u_{90\%}\approx 0.0001$.}  
\label{fig:fig3} 
\end{figure}

\begin{figure} 
\centerline{\includegraphics[width=0.55\linewidth]{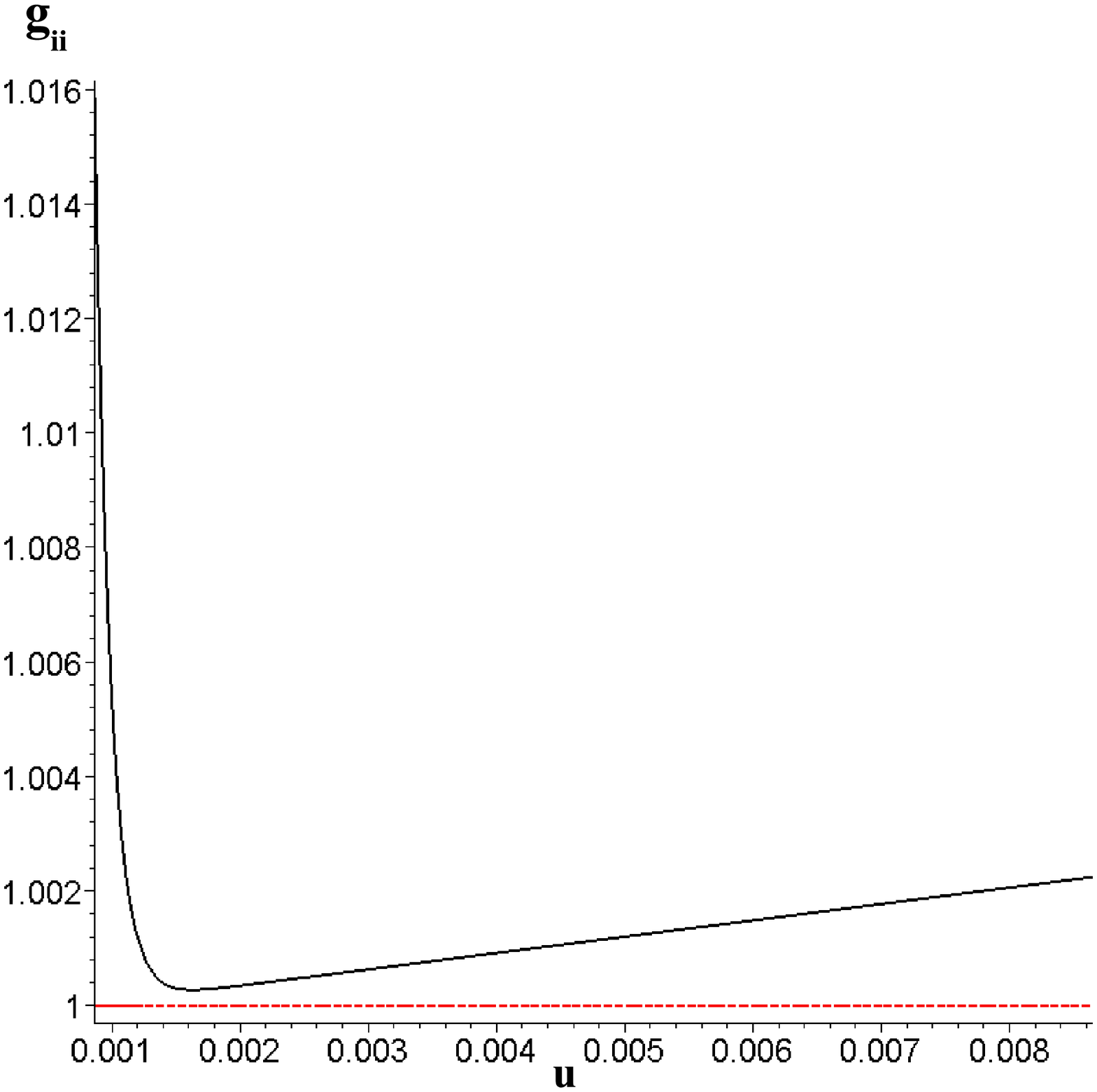}} 
\caption{Plot of the $g_{ii}$ component of the metrics (\ref{eq:metric}). 
The solid black curve represents the numerical solution with $P=a=1000$ and $Q=1$,
and the dashed red curve the analytical solution with $P=0$, $a=1000$ and $Q=1$.
Here $u_{90\%}\approx 0.0001$.}  
\label{fig:fig4} 
\end{figure}

\begin{figure} 
\centerline{\includegraphics[width=0.55\linewidth]{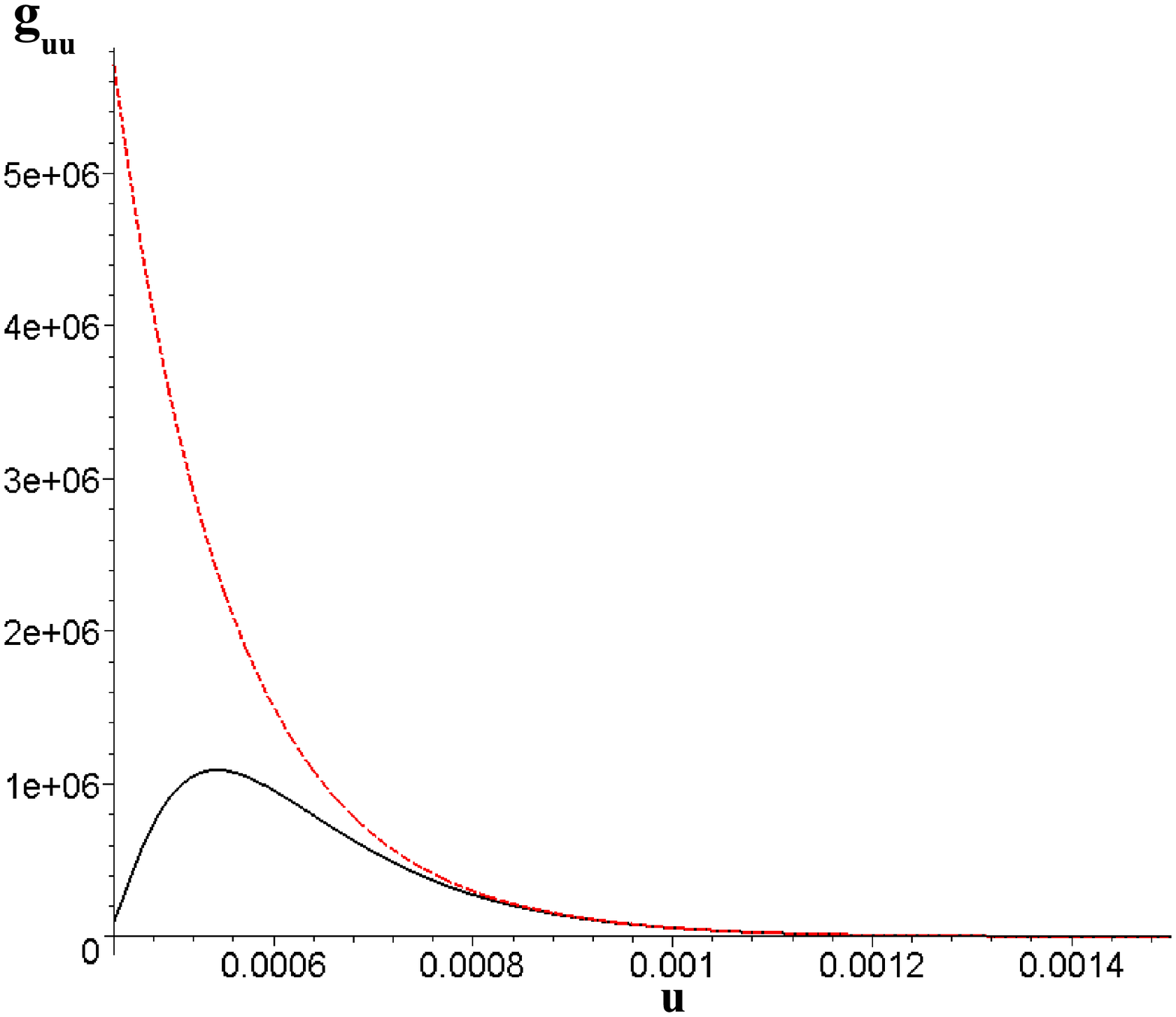}} 
\caption{Plot of the $g_{uu}$ component of the metrics (\ref{eq:metric}). 
The solid black curve represents the numerical solution with $P=a=1000$ and $Q=1$,
and the dashed red curve the analytical solution with $P=0$, $a=1000$ and $Q=1$.
Here $u_{90\%}\approx 0.0001$.}  
\label{fig:fig5} 
\end{figure}

\begin{figure} 
\centerline{\includegraphics[width=0.55\linewidth]{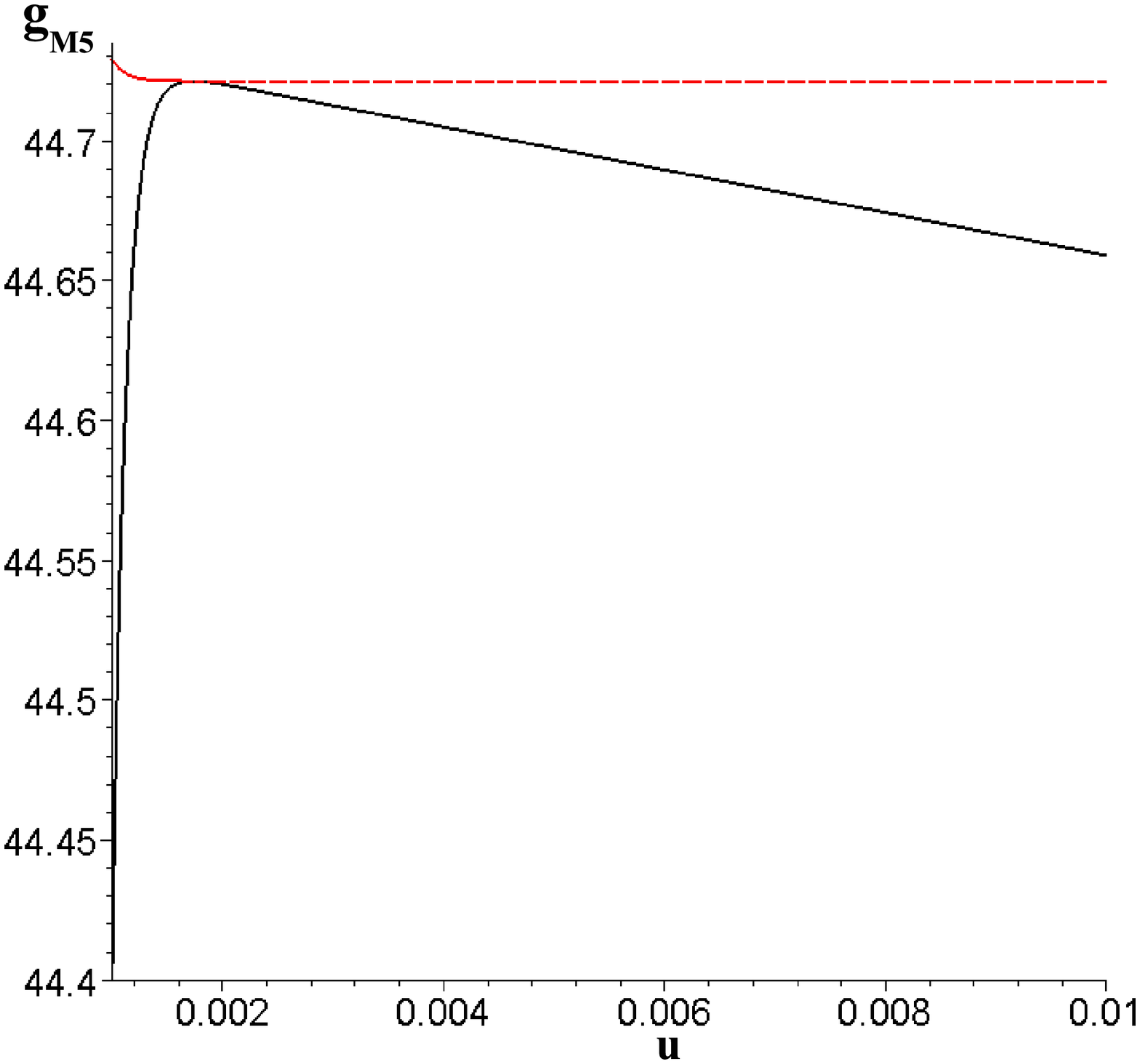}} 
\caption{Plot of the $g_{M5}$ component of the metrics (\ref{eq:metric}). 
The solid black curve represents the numerical solution with $P=a=1000$ and $Q=1$,
and the dashed red curve the analytical solution with $P=0$, $a=1000$ and $Q=1$.
Here $u_{90\%}\approx 0.0001$}  
\label{fig:fig6} 
\end{figure}

In Figs. \ref{fig:fig3}, \ref{fig:fig4}, \ref{fig:fig5} and \ref{fig:fig6}
solid black curves represent the metrics functions obtained numerically
for $P=a=1000$ and $Q=1$. We have chosen these values of the parameters
to make sure the solution is quite different from the D3 or KT solutions.
In these same figures,
the corresponding nonextremal D3 solution,
 i.e., that with $P=0$,
have been drawn with red dashed curves.
As it can be observed, the numerical solution differs from the D3 one.
The difference is more pronounced for $u\to 0$,
but in Figs. \ref{fig:fig4} and \ref{fig:fig6} we note that 
even close to the horizon there is some significant divergence between both solutions.
Here we need to take into account that for these values of the parameters 
$u_{90\%}\approx 0.0001$.

\begin{figure} 
\centerline{\includegraphics[width=0.55\linewidth]{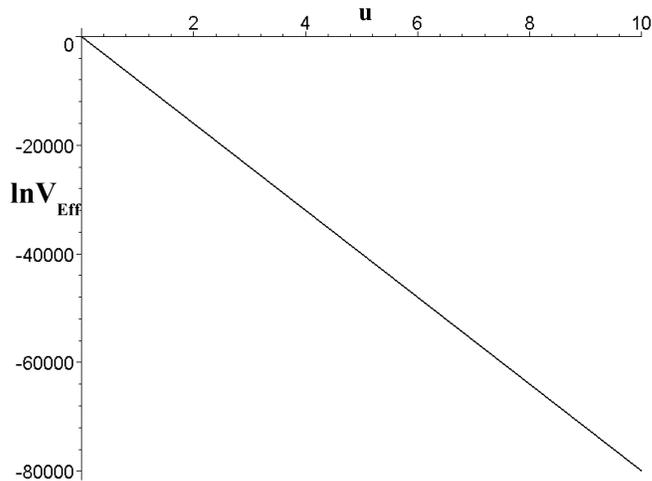}} 
\caption{Plot of the logarithm of the effective potential. Here $u_{90\%}\approx 0.0001$.}  
\label{fig:fig7} 
\end{figure}

\begin{figure} 
\centerline{\includegraphics[width=0.55\linewidth]{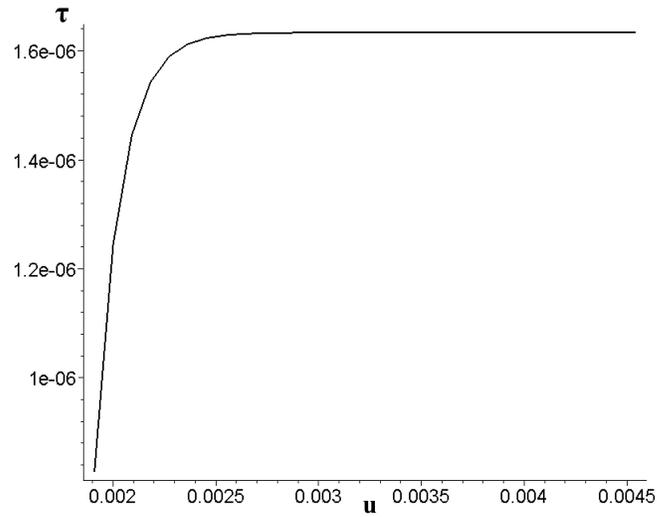}} 
\caption{Plot of the affine parameter. Here $u_{90\%}\approx 0.0001$.}  
\label{fig:fig8} 
\end{figure}

\begin{figure} 
\centerline{\includegraphics[width=0.55\linewidth]{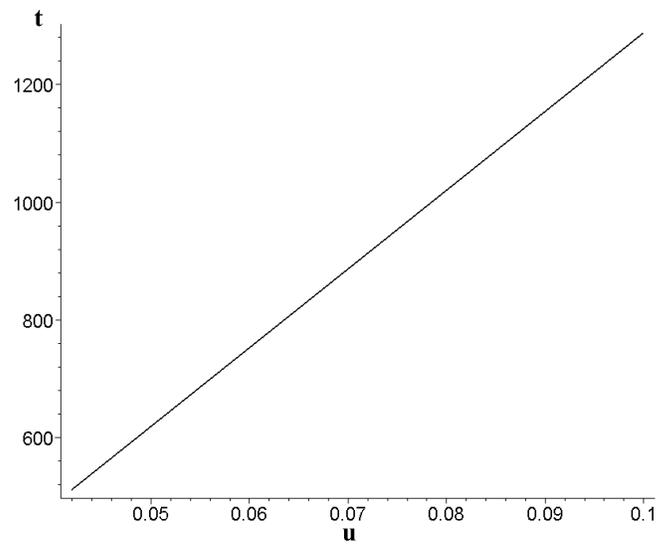}} 
\caption{Plot of the asymptotic time. Here $u_{90\%}\approx 0.0001$.}  
\label{fig:fig9} 
\end{figure}

\begin{figure} 
\centerline{\includegraphics[width=0.55\linewidth]{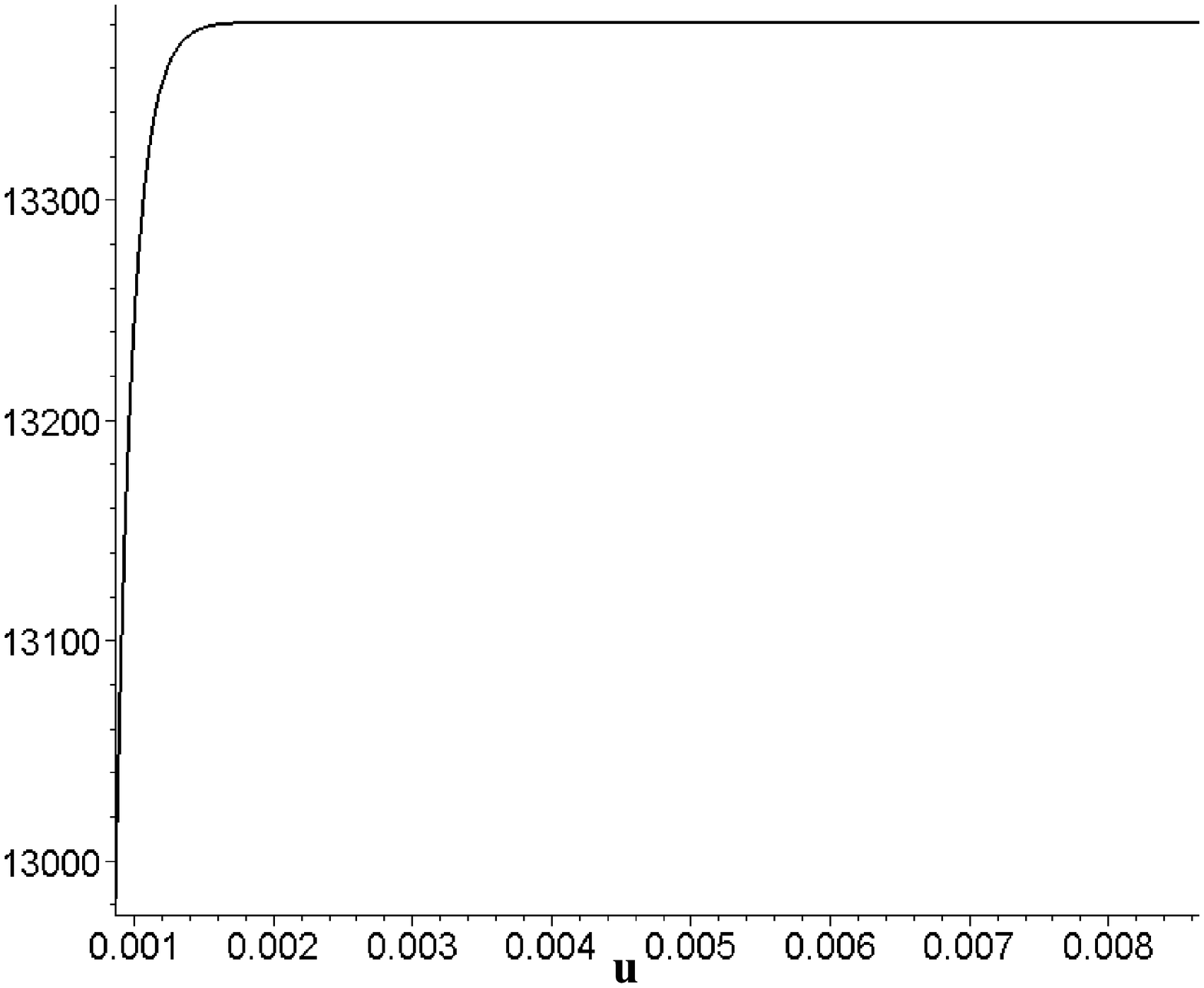}} 
\caption{Plot of $\exp\left(-2z+3x+5y\right)$.
In the limit $u\to \infty$ this quantity is proportional to the horizon area.
Here $u_{90\%}\approx 0.0001$.}  
\label{fig:fig10} 
\end{figure}

Nevertheless, the numerical results certainly indicate the existence of a black hole.
To show that, in Figs. \ref{fig:fig7}, \ref{fig:fig8}, \ref{fig:fig9} and \ref{fig:fig10}
we present, respectively, 
 the natural logarithm of the effective potential $V_{Eff}\equiv e^{2z-6x}$
entering equation (\ref{eq:VEff}), 
the proper and asymptotic times as given by equations (\ref{eq:affineT}) and (\ref{eq:asymptT}),
and quantity $e^{-2z+3x+5y}$ which, 
as it was mentioned in section \ref{sssec:nonextremald3}, 
is proportional to the black hole horizon area when $u\to \infty$. 

Indeed, the following key properties are observed: 
absence of a potential barrier in the effective potential,
the affine parameter converges to a finite value while the test particle approaches the horizon,
simultaneously the asymptotic time diverges,
and the corresponding horizon area is finite.
Thus, we conclude that this solution represents a regular black hole for $u\to \infty$.

\begin{figure} 
\centerline{\includegraphics[width=0.55\linewidth]{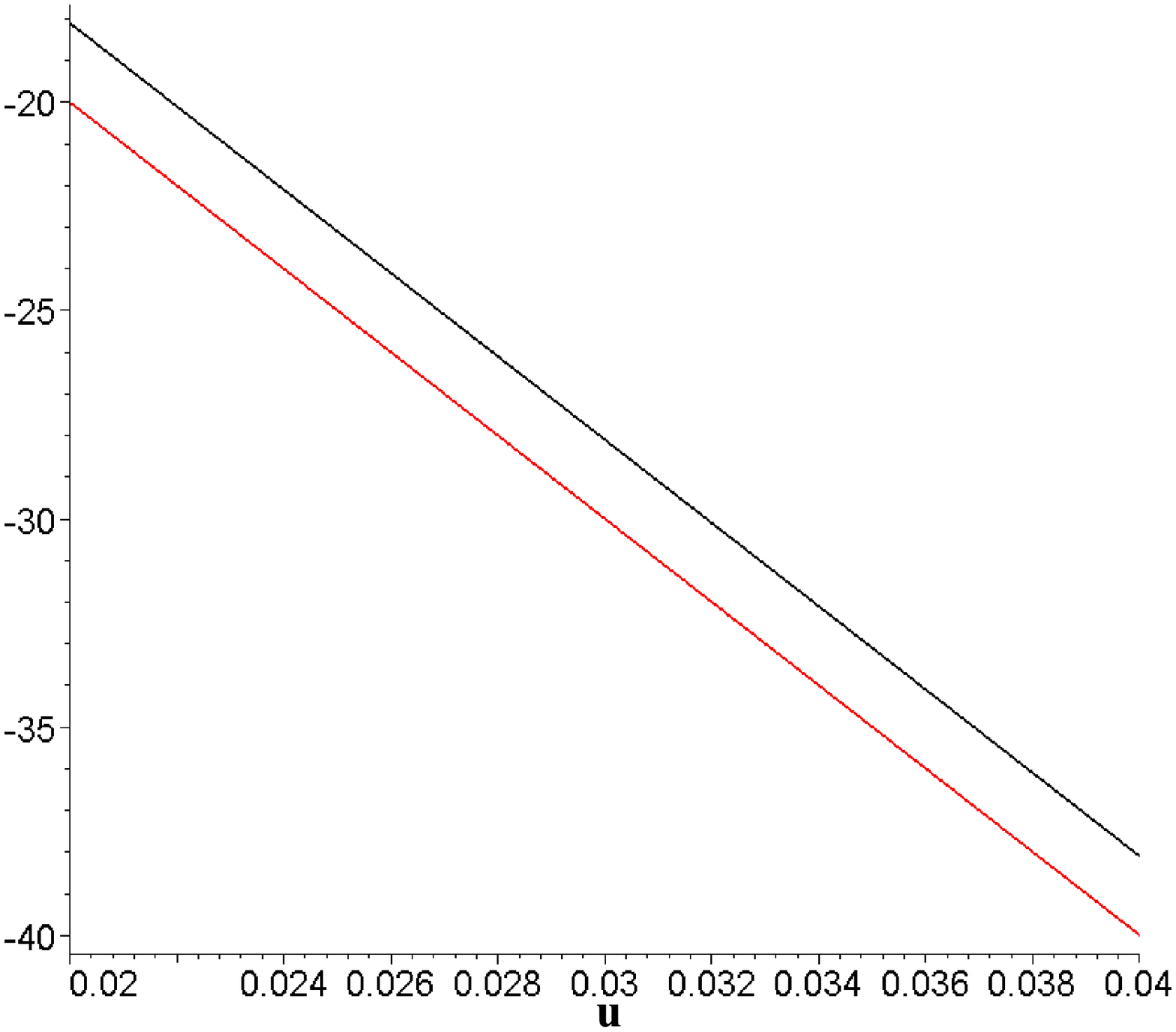}} 
\caption{Plot of the fields $y$ (black) and $z$ (red).
Here $u_{90\%}\approx 0.0001$.}  
\label{fig:fig11} 
\end{figure}

As discussed earlier in subsection \ref{ssec:unumbers}, 
we can assess the quality of these results by analyzing the asymptotic behavior of fields $y$ and $z$,
which are plotted in figure \ref{fig:fig11}.
It is observed that both fields behave as linear functions of $u$. The linear behavior is required from 
the existence and regularity of the horizon, see discussion around (\ref{eq:yzasym}). 
As a matter of fact, a least-squares fit to a polynomial of this numerical output yields,
\begin{eqnarray}
y_{\infty}=1.90027- 999.942947\,u+ 0.0000002\,{u}^{2}+0.00000002\,{u}^{3}+ 0.000000002\,{u}^{4}\, ,\\
z_{\infty}=- 0.000113- 999.857345\,u+ 0.0000004\,{u}^{2}+ 0.0000001\,{u}^{3}+ 0.00000001\,{u}^{4}\, .
\end{eqnarray}
We see that higher order coefficients are negligible.
Using that here $a=1000$, we find that $\alpha=-0.999857$ and $\beta=-0.999943$,
yielding $-2\alpha + 5\beta = -3.00000004$.
This is indeed, a strong result supporting the reliability of the numerical solution,
as well as the corresponding asymptotical analysis presented in 
the last part of subsection \ref{sssec:nonextremald3}.

According with expression (\ref{eq:BHT}) and the above results, 
the temperature of this black hole is equal to $0.035695$ in the appropriate units.

\subsection{Asymptotical behavior at infinity, $u=0$}\label{ssec:u0}

While studying the asymptotic behavior of the numerical solution at infinity,
 $u \to 0$,
we have found that typically the $z$-field diverges at some $u=u_{sing}$,
while the remaining fields seem to be analytical at that point.
Numerically, the value of $u_{sing}$ can be positive or negative.
If $u_{sing}<0$, it means that the corresponding space-time is complete,
and it is asymptotically flat.
If $u_{sing}>0$, the corresponding space-time 
is singular in the sense that it collapses at a finite value of the radial coordinate. 
In principle, there exists also the possibility of $u_{sing}=0$.
In this last case, the space-time is complete and it has a potential wall at infinity,
like in AdS.
Independently of the value of $u_{sing}$,
the qualitative behavior of the solutions is essentially the same.
As we will see soon,
obtaining one case or the other
depends mainly on the value of $a$,
as well as, on the boundary conditions.

\begin{figure} 
\centerline{\includegraphics[width=0.55\linewidth]{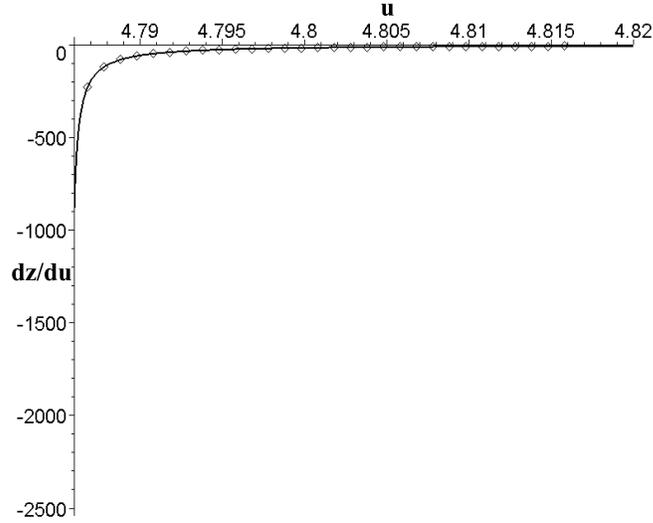}} 
\caption{Understanding the singular behavior near $u=u_{sing}$.
The diamonds represent the numerical output of dz/du,
while the curve is the plot of the function given by equation (\ref{eq:dzAnti}).
Here $u_{90\%}\approx 173$ and $u_{sing}\approx 4.78571$. }  
\label{fig:fig12} 
\end{figure}

\begin{figure} 
\centerline{\includegraphics[width=0.55\linewidth]{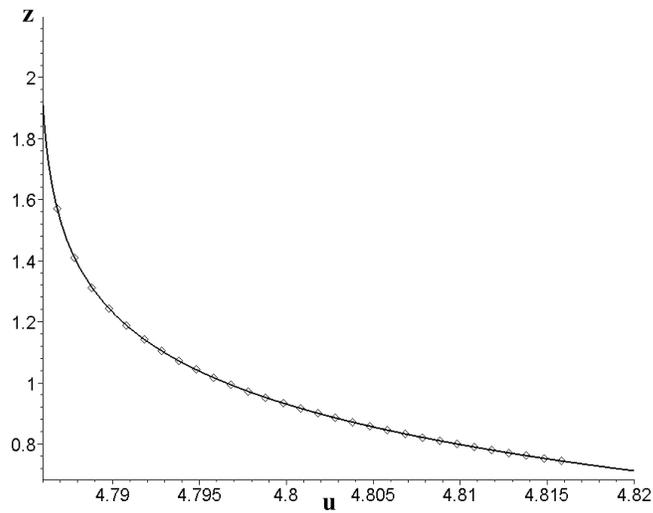}} 
\caption{Understanding the singular behavior near $u=u_{sing}$.
The diamonds represent the numerical output of z,
while the curve is the plot of the function given by equation (\ref{eq:zAnti}).
Here $u_{90\%}\approx 173$, $u_{sing}\approx 4.78571$ and $A\approx 1.6$.}  
\label{fig:fig13} 
\end{figure}

The interesting result here is that
this singularity has a lot in common with the one in the KT solution.
If we assume that,
\begin{equation}
    z\stackrel{\,\,\,\,\,\,\,\,u\rightarrow u_{sing}}{=}
    -\frac 14 \ln\left[A u \ln\left(\frac u{u_{sing}}\right)\right]\, ,
\label{eq:zAnti}        
\end{equation}
where $A$ is a constant,
then the corresponding derivative,
\begin{equation}
  \frac{dz}{du}\stackrel{\,\,\,\,\,\,\,\,u\rightarrow u_{sing}}{=}
    -\frac 14 \frac{\ln\left(\frac u{u_{sing}}\right) + 1}{u\ln\left(\frac u{u_{sing}}\right)}  \, ,
\label{eq:dzAnti}   
\end{equation}
depends only on $u_{sing}$,
which is given by the numerical method.
In figure \ref{fig:fig12} 
the diamonds represent the numerical solution for $dz/du$ with $P=1$, $Q=1$, $a=0.01$. 
For this case $u_{90\%}\approx 173$ and we found $u_{sing}\approx 4.78571$. 
As it can be noted, 
the behavior behind the numbers is very accurately reproduced by
the solid curve corresponding to the plot of the function given by equation (\ref{eq:dzAnti}).
The corresponding behavior for $z(u)$ is shown in figure \ref{fig:fig13}.
The main difference between these expressions and the KT solution is that in the latter 
the range of $u$ is $0 < u < u_{sing}$,
while here $\infty> u >u_{sing}$.
This requires a change in the sign of $A$ and, 
indeed,
we have found that here $A\approx 1.6 > 0$, 
while for KT, $A\equiv -P^2/2$.
To confirm these findings,
we fitted the solutions for $y$ and $f$ 
(and their derivatives),
using respectively the following expressions,
\begin{eqnarray}
\label{eq:yAnti}
y &\approx&
    -\frac 14 \ln|4u|+{\it y\_0}+{\it y\_1} \left( u-u_{sing}\right) +{\it y\_2} \left( u-u_{sing} \right) ^{2}\, ,\\
\label{eq:fAnti}    
f &\approx&
    -\frac1{2P} \left[ B\ln  \left( \frac {u \textbf{\textit{e}}}{u_{sing}} \right) + Q \right] +
{\it f\_0}+{\it f\_1} \left( u-u_{sing}\right) +{\it f\_2} \left( u-u_{sing} \right)^2\, ,
\end{eqnarray}
where $\textbf{\textit{e}}$ is the Euler number.
Solving a least-squares problem for the coefficients,
we obtained
${\it y\_0}=-0.041239,{\it y\_1}=0.07182457,{\it y\_2}=-0.128777,
B=0.905268496,{\it f\_0}=1.274356,{\it f\_1}=0.285765,{\it f\_2}=-0.05121489$.
The fits are amazingly good, with $\chi^2<10^{-10}$ for both cases.
Since, these variables are analytical at $u_{sing}$,
we tried a direct fit with Taylor expansions for them,
as well as for the remaining fields. 
The solution of the corresponding least-squares problems yielded,
\begin{eqnarray}
y(u)&\approx& -0.783602 - 0.042953(u-u_{sing}) + 0.003493(u-u_{sing})^2 \nonumber\\
  &+& 0.001668(u-u_{sing})^3 - 0.115651(u-u_{sing})^4 +3.027465(u-u_{sing})^5 ,\\
f(u)&\approx& 0.34785 - 0.061501(u-u_{sing}) + 0.008764(u-u_{sing})^2 \nonumber\\
  &-& 0.002828(u-u_{sing})^3 + 0.099713(u-u_{sing})^4 - 2.972385(u-u_{sing})^5,\\       
\Phi(u)&\approx& 0.231423 - 0.027829(u-u_{sing}) + 0.002474(u-u_{sing})^2 \nonumber\\
  &-& 0.002109(u-u_{sing})^3 + 0.120818(u-u_{sing})^4 - 3.620755(u-u_{sing})^5,\\
w(u)&\approx& 0.028572 - 0.004287(u-u_{sing}) + 0.000513(u-u_{sing})^2 \nonumber\\
  &-& 0.000065(u-u_{sing})^3 + 0.000009(u-u_{sing})^4 - 10^{-6}(u-u_{sing})^5.
\end{eqnarray}
We recall that these results were checked by also fitting the corresponding derivatives. 
Expanding the non-polynomial part of the expressions (\ref{eq:yAnti}) and (\ref{eq:fAnti})
it is not difficult to test the direct correspondence 
between both approximations of the behavior of $y$ and $f$ when $u\to u_{sing}$.    

Substituting in system (\ref{eq:sys}) the ansatz (\ref{eq:zAnti}) for the $z$-field 
and Taylor-like parametrizations for the remaining fields near $u_{sing}$,
it can be proved that they provide an asymptotic solution
if the following condition is satisfied,
    \[
f_1=\pm P\rm{e}^{\left(\Phi_0+4y_0+4w_0\right)}\, . 
\]
We can check if this analytic result is verified numerically. Indeed, using the numbers in the 
Taylor expansion of the fields given above from the numerical analysis, 
and $P=1$,
we can verify that
    \[
- 0.061501 \approx f_1  = - \rm{e}^{0.231423+4(-0.783602)+4(0.028572)}\approx -0.061501\, . 
\]
Thus the analytic condition is indeed fulfilled by the numerical analysis.
As a matter of fact, 
we have found that this condition is satisfied by all the solutions we analyzed.
For instance, the solution with $P=Q=1$ and $a=0.1$ is regular at the origin,
because $u_{sing}=-0.056996$ which corresponds to an asymptotically flat solution . 
For this case we also verify that,
    \[
-37.008 \approx f_1 = - \rm{e}^{0.36724+4(0.693744)+4(0.117277)}\approx -37.015\, .   
\]
This is a very strong validation of our numerical results 
as well as of the analytical solution for the $u\to u_{sing}$ asymptotics given by expressions 
(\ref{eq:zAnti}), (\ref{eq:yAnti}) and (\ref{eq:fAnti}).

From the point of view of field theory the more interesting solution is the one with $u_{sing}=0$.
We have found that the position of the singularity near the origin depends mainly on the values of $a$
and of the boundary conditions.
The dependence on $a$ seems to be described by a convex function $u_{sing}(a)$
having a minimum at some $a=a_c$. 
This minimum will be positive or negative depending on the boundary conditions.
There is a unique set of boundary conditions that lead to $u_{sing}(a_c)=0$.
The main difficulty to find such a solution 
is that, because of the finitness of the integration step, 
falling into a point with $u_{sing}<10^{-10}$ 
is very difficult
when the integration starts far from that value.
We should also recall that ours is a nonlinear system, 
therefore it is sensitive to changes of the parameters
and strongly sensitive to changes of the boundary conditions.
There is also a restriction on varying the boundary conditions 
coming from the necessity of preserving the black hole solution at $u\to \infty$. 
Nevertheless,
we were able to obtain solutions near enough to the origin
(for instance, with $|u_{sing}| = \textsl{O}(10^{-6})$)
and independently of the sign 
we have always found that the whole analysis in this section applies,
so there is no particular reason to expect the situation to be different for $u_{sing}=0$.

To finish this section we would like to remark that
we observed that after reaching the minimum, 
$u_{sing}$ gets larger while still increasing $a$.
On the other hand, $u_{90\%}$ always decreases while increasing $a$.
So, there exists a not necessarily high value of $a$ such that $u_{sing}=u_{90\%}$.
As a matter of fact, 
that is true for any distance until the horizon as estimated by expresion (\ref{eq:u90}).
It implies that 
there seems to always exist a maximum value of $a$ such that there is no solution outside the black hole horizon.

\subsection{Area and temperature dependence on the parameters}\label{ssec:Thermo}

In this paper we will not attempt a full analysis of the dependence of the solution on the 
three parameters $a$, $P$ and $Q$,
as well as on the boundary conditions. We concentrate on explicitly constructing  numerical solutions. 
Nevertheless, we would like to present some observations that show  
 the power of the numerical method to understand the physics behind the more general solutions.
The naive intuition is that the properties of the solution are controlled by parameters like
\be 
\frac{P^2}{Q}, \quad \frac{a}{Q}, \quad \frac{a}{P^2}. 
\ee 
This was however, not supported by the numerical analysis. 

As we already mentioned fixing $P=0$ always yields the nonextremal D3 with regular horizon. 
Increasing $P$ from zero the solutions differs very slowly from the D3 solution. 
For values lower than $P=10^{-6}$, 
the existing black hole solutions cannot be practically distinguished from the D3 one.
Nevertheless, it must be noted that some intervals of $P$ arise where black hole solutions cease to exist.
For instance, 
we failed to find a black hole
for $Q=10^{-4}$ and $a=0.1$ and $P=10^{-8}$.
Above some value of $P$, 
the horizon area does not seem to converge to a finite value.  
And this is not necessarily a very high value of P. 
For instance, 
for $Q=10^{-4}$ and $a=0.1$, as above, and $P\geq 5\times10^{-2}$,
we found nothing but degenerate black hole solutions. 
So, we are left here with the range $P\in[10^{-6},5\times10^{-2})$,
where the dependence on this parameter is very weak. 
In principle, as one increases the value of $P$ the temperature also increases. 
The horizon area has the same behavior.
On the other asymptotic regime of $u$, for small u, 
the larger $P$, 
the lower the value of $u_{sing}$ seems to be. 
However, this behavior is altered near those islands where solutions with regular horizons do not exist.

\begin{figure} 
\centerline{\includegraphics[width=0.55\linewidth]{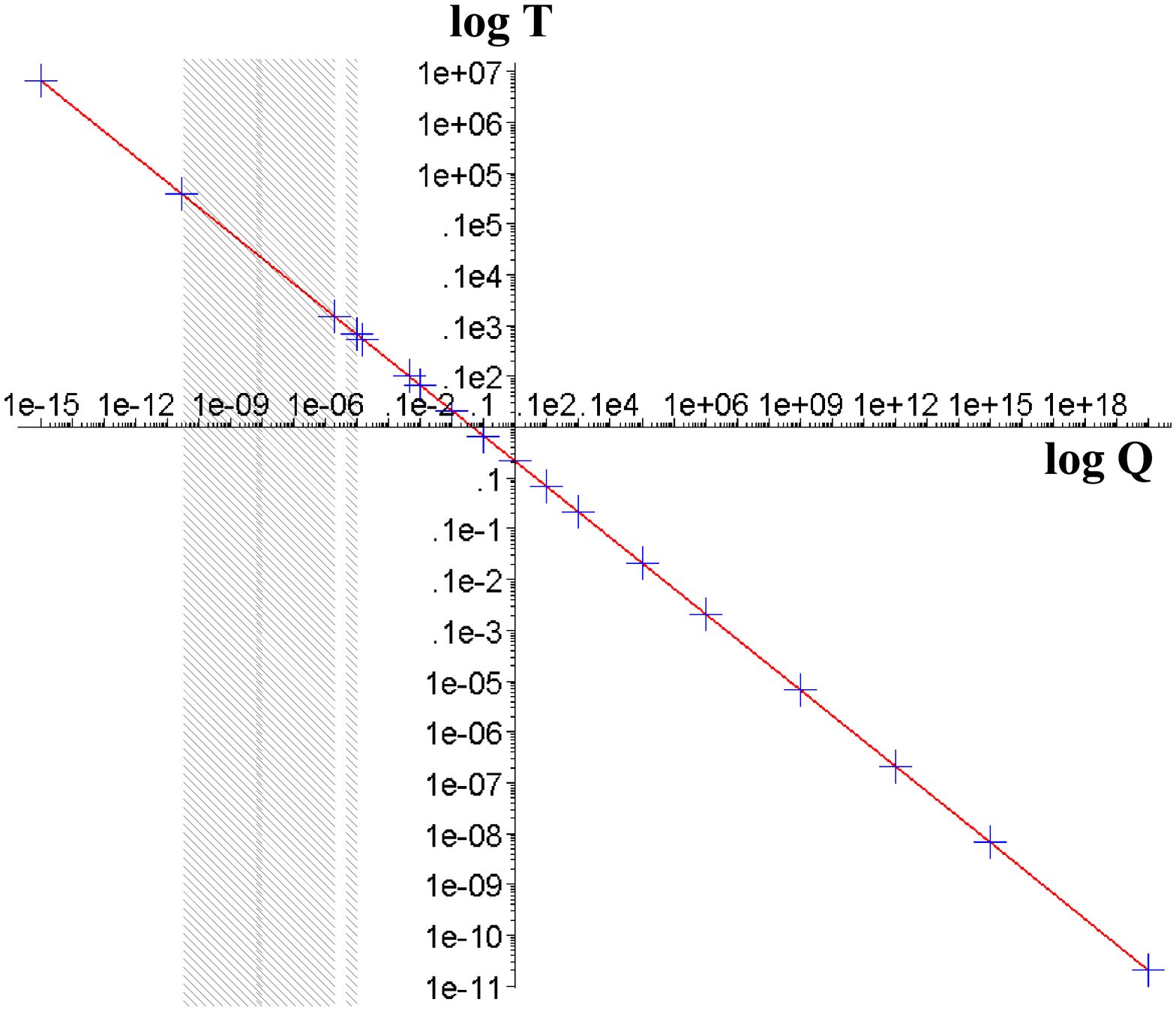}} 
\caption{The dependence of the black hole temperature on $Q$. 
The shaded regions mark some intervals of $Q$ where no black hole solution seems to exist.}  
\label{fig:fig14} 
\end{figure}

The dependence on $Q$ appears to be the opposite of what we observed for $P$.
For instance, with $P=10^{-7}$ and $a=0.1$, 
for $Q<10^{-10}$ the horizon area does not converge to a finite value. 
Increasing $Q$ up to $4.4\times 10^{-4}$ 
some intervals arise where black hole solutions exist and others where they do not.
If a black hole solution exists, 
then for all value of $Q$ it has a finite horizon area. 
For $4.4\times 10^{-4}\leq Q<10^{20}$ all the solutions we found were regular black holes.
As in the D3 solution,
$Q$ does not seem to affect the asymptotic behavior of $y$ when $u$ goes to infinity, 
but affects the asymptotics of $z$.
As it is shown in figure \ref{fig:fig14}, 
$Q$ is inversely proportional to the temperature. The horizon area has the inverse behavior.
This kind of dependence are proper also of the D3 solution.
The value of $u_{sing}$ does not seem to depend on $Q$.

\begin{figure} 
\centerline{\includegraphics[width=0.55\linewidth]{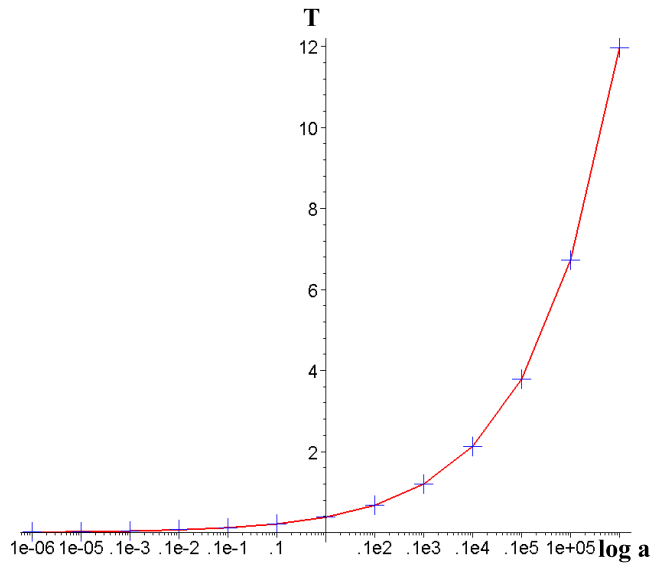}} 
\caption{The dependence of the black hole temperature on $a$.}  
\label{fig:fig15} 
\end{figure}

The first thing we note while varying $a$ is that, 
given a combination of $P$ and $Q$, 
no value of $a$ exist (except $a=0$) that could change one of the following results: 
no black hole, singular black hole, regular black hole.
The larger $a$, the higher the temperature,
which is also the kind of dependence observed in the D3 solution.
In figure \ref{fig:fig15}
we present the result for $P=10^{-7}$ and $Q=1$. 
The horizon area presents a similar behavior. 
Figure \ref{fig:fig15} suggests that the horizon temperature and area blow up at some finite value of $a$.
That confirms the findings described at the end of section \ref{ssec:u0},
where it was discussed the dependence on $a$ of the critical value $u_{sing}$.
It means that there seems to exists a maximal value of $a$ 
above which no black hole with finite temperature or area can exist.

\section{Conclusions}\label{conclusions}

In this paper we have solved numerically the IIB supergravity equations of motion in the 
presence of varying flux. This class of solutions is controlled by three parameters loosely 
identified as $Q$ with the five-form flux, $P$ with the three-form flux and $a$ with nonextremality. 

We established that choosing the right boundary conditions, and for a wide range of values of the 
parameters $P$, $Q$ and $a$,  we can obtain solutions representing black holes with 
finite horizon area. The criteria to assert the existence of the black hole are: no barrier of the 
effective gravitational potential, convergence of the affine parameter, divergence of the asymptotic time 
and finiteness of the horizon area. In this regime the obtained solutions resemble but are 
not identical to the standard non-extremal D3 solution.  
 
We also detected some curious behavior which deserves a more systematic study. Let us summarize how the 
horizon temperature and its area depend on the parameters of the solution. 
In the parameter space $(P,Q)$ there are islands of values which do not lead to black hole solution.
This particular result weakly depends on the values of $a$.
When a black hole exist, the temperature and area of the horizon seem 
to be proportional to the value of  $P$. 
Above some value of $P$, the horizon area ceases to converge to a finite value.
The dependence of the horizon temperature and the area on $Q$ is very much like that in the D3 solution, i.e,
the higher $Q$, the higher the horizon area but the lower its temperature.
Curiously, below some threshold value of $Q$ the horizon area is no longer finite.
Finally, the dependence of the thermodynamical quantities on $a$ is quite interesting.
It resembles the situation in the D3 solution, the area  and the temperature are both proportional to $a$.
However, here our results indicate the possibility of the existence
of a maximal value of $a$ above which finite-temperature black hole can not exist.

In the region asymptotically away from the horizon we were able to obtain a very accurate description 
based on a combination of analytic and numerical methods. We found that the solution has qualitatively a 
KT-like asymptotic behavior but with modification allowing the solutions to exist for $u>u_{sing}$. 
Depending on whether $u_{sing}$ is negative or positive we have an asymptotically flat 
or a collapsing space-time. There seems to exist a solution with $u_{sing}=0$ which is complete and 
behaves asymptotically as the AdS space-time.

Let us end our remarks about the solution by presenting the dependence of  $u_{sing}$ on the parameters of the 
solution. While increasing $a$, $u_{sing}$ initially decreases, reaches a minimal value and then starts to increase. 
This implies the existence of a maximal value of $a$ above which no black hole with finite temperature 
or area can exist. The dependence of $u_{sing}$ on the remaining parameters is more complex and deserves further study. 
Though weakly, the value of $u_{sing}$ seems to be inversely proportional to $P$ and almost independent of $Q$.

In this paper we have focused on constructing the solution. There are many ideas that would be interesting to 
explore on this supergravity background. 
An important venue that we plan to return to \cite{pterr} is the possibility of a 
Hawking-Page phase transition between the 
solution constructed here and the Klebanov-Strassler background. Some of the various  questions 
we plan to address in future publications  have 
been addressed recently in simpler supergravity backgrounds \cite{cobitemp,temp1,temp2,temp3,temp4}.

\section*{Acknowledgments}

We thank Don Marolf and Cobi Sonnenschein for interesting comments. 
Part of this work was carried out while LAPZ visited the KITP and we acknowledge its hospitality. 
This work is  partially supported by Department of Energy under 
grant DE-FG02-95ER40899 to the University of Michigan 
and by grant CNPq/CLAF-150548/2004-4.

\appendix

\section{Geodesic analysis of some typical solutions}\label{geodesics}
In this appendix we explicitly discuss the geodesic structure of some known solutions. The 
idea is to develop the necessary intuition to interpret the numerical results. We start by 
considering geodesic motion in the general metric  (\ref{eq:metric}). 
The effective potential for radial motion is given by 
\be
\call=-e^{2z-6x}\dot{t}^2 + e^{-2z+10y}\dot{u}^2.
\ee
Normalizing the effective Lagrangian to be $\call =-1$ and using the fact that the Lagrangian 
does not explicitly depends on time we find an equation for the radial coordinate:
\be
\dot{u}^2 +e^{2z-10y}=E^2 \, e^{6x-10y}.
\ee
This equation does not readily allows an interpretation in terms of classical motion of a particle 
with energy $E$ in a potential. For this aim, we find it useful to define a new variable $v$ given by
$ dv= e^{-3x+5y}du$. 
The geodesic equation then becomes
\be
\eqlabel{eq:potential}
\dot{v}^2+e^{2z-6x}=E^2,
\ee
where the functions $x,y$ and $z$ are now viewed as functions of $v$.  
In this presentation we can consider 
the function $e^{2z-6x}$ as an effective gravitational potential describing the classical motion of a particle. 

\subsection{AdS space}
For AdS space in Poincare coordinates which are the natural coordinate for the D3 brane we find 
\be
\call = -\frac{r^2}{L^2}\dot{t}^2 + 
\frac{L^2 \dot{r}^2}{r^2}.
\ee
The equation for radial motion as a function of the proper time $\tau$ is: 
\be
\dot{r}^2+\frac{r^2}{L^2} =E^2.
\ee
The corresponding potential is $V(r)=r^2/L^2$. 
This equation can be easily recognized as the harmonic oscillator whose solution is
\be
r(\tau)=EL \, \sin(\frac{1}{L}\, \tau +\phi),
\ee
where $\phi$ is a phase that determines the initial condition for the motion of the particle. 

Our main message is that ``physical'' experiments in AdS are quite different from Schwarzschild. As we can see from the 
potential, AdS functions acts as a box and the experiments we can perform involve sending a particle from the 
interior of AdS towards the boundary $r=\infty$. Such particle explores the boundary and returns back having 
reached its maximum radius $r_{max}=EL$. 

\subsection{Nonextremal D3 branes}

For the nonextremal D3 brane in the standard parametrization the equation for massive geodesic is 
\be
\dot{r}^2+\frac{r^2}{L^2}-\frac{r_0^4}{L^2 \, r^2} =E^2.
\ee
Note that taking $r_0=0$ yields an equation which we recognize as the harmonic oscillator describing motion of 
massive particles in AdS. More generally, we can view this problem as a classical mechanical problem in a potential
given by 
\be
V(r)=\frac{r^2}{L^2}-\frac{r_0^4}{L^2\, r^2}.
\ee
For large $r$ we have the typical harmonic potential of AdS but for $r\to r_0$ the 
potential goes to zero.

The ``physical'' experiment here is slightly different. We have motion between $r=r_0$ and some $r_{max}$ given by:
\be
(EL)^2=r_{max}^2+ \frac{r_0^4}{r_{max}^2},
\ee
Note that $r_{max}$ in this case is larger than $EL$ which corresponds to AdS with no black hole. Namely, 
\be
r_{max}^2=\frac{(EL)^2}{2}\left(1+\sqrt{1+\frac{4r_0^4}{(EL)^4}}\right)
\ee

The massive geodesic in the $\rho$ coordinates has
\bea
\tau_{massive}&=&\int \frac{d\r}{\sqrt{E^2 -\frac{\r^2}{L^2}\left(1-\frac{\rho_0^4}{\rho^4}\right)}} \nonumber \\
t&=& \int \frac{d\r}{\sqrt{E^2 -\frac{\r^2}{L^2}\left(1-\frac{\rho_0^4}{\rho^4}\right)}}
\frac{E\, L^2}{r^2\left(1-\frac{\rho_0^4}{\rho^4}\right)},
\eea

\subsection{The structure of the KT solution}
Let us verify that the KT solution has a behavior similar to the negative mass Schwarzschild solution and that
it can be detected by the effective potential (\ref{eq:potential}). 
Consider a solution of KT type with metric given by
\be
ds^2 = h^{-1/2}dt^2 + h^{1/2}dr^2 + \ldots
\ee
where the warp factor is of the form
\be
h=\frac{L^4}{r^4}(1+P\ln\frac{r}{r_0}),
\ee
A massive geodesic follows an equation of the form:
\be
\dot{r}^2 + \frac{r^2}{L^2\sqrt{1+P\ln\frac{r}{r_0}}}=E^2,
\ee
Let us, once again, consider the natural limits. For large values of $r$ the potential is mainly harmonic 
$V(r)\sim r^2/ \ln^{1/2}(r)$. This behavior is typical of asymptotically AdS spaces as we saw at the beginning 
of the section. 
For the purpose of a horizon we are interested in whether or not the potential develops a wall for small values 
of $r$. Indeed, it is clear that the denominator of the potential is zero, that is, near 
\be
r_{sing}=r_0\exp(-1/P).
\ee
Around this point the potential goes to infinity and 
the spacetime behaves like in the neighborhood of a negative-mass Schwarzschild black hole.  Before this point there is a minimum of 
the potential located at 
\be
r_{min}=r_{0}\exp(\frac{P-4}{4P})=r_{sing} e^{1/4}.
\ee


\end{document}